\def\cdtx{{\scriptstyle\bullet}}
\def\cdt{{\scriptscriptstyle\bullet}}
\def\txi#1{\widetilde{\bxi}\phantom{x}\hspace{-1.2ex}^{#1}}
\def\phan{\left.\phantom{|^|_|}\!\!\!}
\def\begB{\setcounter{qqq}{1}\renewcommand{\theequation}{\arabic{equation}\alph{qqq}}\begin{equation}}
\def\endB#1#2{\label{#1}\end{equation}\vspace*{-\baselineskip}\phantom{ }\newcounter{#2}\setcounter{#2}{\value{equation}}}
\def\begI{\addtocounter{equation}{-1}\addtocounter{qqq}{1}\begin{equation}}
\def\endI#1{\label{#1}\end{equation}}
\def\endL#1{\label{#1}\end{equation}\vspace*{-\baselineskip}\renewcommand{\theequation}{\arabic{equation}}\break}
\def\mi{\vspace*{2mm}\noindent}
\def\lbd{\{\hskip-4pt\{}
\def\rbd{\}\hskip-4pt\}}
\def\ladb{\left\langle\phantom{|^|_|}\hskip-10pt\right\langle}
\def\radb{\left\rangle\phantom{|^|_|}\hskip-10pt\right\rangle}
\def\lbdb{\left\{\phantom{|^|_|}\hskip-11pt\right\{}
\def\rbdb{\left\}\phantom{|^|_|}\hskip-11pt\right\}}
\def\lpar{\left(\phantom{|^|_|}\hskip-9pt\right.}
\def\rpar{\left.\phantom{|^|_|}\hskip-9pt\right)}
\def\lad{\langle\hskip-3pt\langle}
\def\rad{\rangle\hskip-3pt\rangle}
\def\lb{\{}
\def\rb{\}}
\def\la{\langle}
\def\ra{\rangle}
\def\bxi{\mbox{\boldmath$\xi$\unboldmath}}
\def\bal{\mbox{\boldmath$\alpha$\unboldmath}}
\def\bbeta{\mbox{\boldmath$\beta$\unboldmath}}
\def\bOm{{\bf\Omega}}
\def\bom{\mbox{\boldmath$\omega$\unboldmath}}
\def\rf#1{(\ref{#1})}
\def\xrf#1#2{(\ref{#1})--(\ref{#2})}
\documentstyle[12pt]{article}
\begin{document}
\newcounter{qqq}
\oddsidemargin 5mm
\renewcommand\thefootnote{\fnsymbol{footnote}}
\renewcommand\arraystretch{2}
{\center

\mi{\bf Amplitude equations for weakly nonlinear two-scale perturbations\\
of free hydromagnetic convective regimes in a rotating layer}

\mi
VLADISLAV ZHELIGOVSKY\footnote{E-mail: vlad@mitp.ru}

\mi
International Institute of Earthquake Prediction Theory\\
and Mathematical Geophysics\\
84/32 Profsoyuznaya St., 117997 Moscow, Russian Federation\\

\mi
Observatoire de la C\^ote d'Azur, CNRS\\
U.M.R. 6529, BP 4229, 06304 Nice Cedex 4, France\\

\mi
({\it Received 9 June 2008; in final form 27 January 2009})

}

\medskip Weakly nonlinear stability of regimes of free hydromagnetic thermal
convection in a rotating horizontal layer with free electrically conducting
boundaries is considered in the Boussinesq approximation. Perturbations are
supposed to involve large spatial and temporal scales. Applying methods for
homogenisation of parabolic equations, we derive the system of amplitude
equations governing the evolution of perturbations under
the assumption that the $\alpha-$effect is insignificant in the leading order.
The amplitude equations involve the operators of anisotropic combined eddy
diffusivity correction and advection. The system is qualitatively different
from the system of mean-field equations for large-scale perturbations of forced
convective hydromagnetic regimes. It is mixed: equations for the mean magnetic
perturbation are evolutionary, all the rest involve neither time derivatives,
nor the molecular diffusivity operator.

\mi{\it Keywords:} thermal convection, magnetic dynamo,
weakly nonlinear stability, amplitude equations, mean-field equations,
$\alpha-$effect, eddy diffusivity

\mi{\bf 1. Introduction}

\mi
This paper is a continuation of Zheligovsky (2008), further referred to as
Zh08, where the relevance of the problem considered here for geophysical
applications is discussed and the work of other authors in this area is
reviewed in detail.

Our ultimate motivation is the same as in Zh08: we would like to find examples
of ``real'' convective dynamos, whose weakly non-linear perturbations have upon
saturation a complex structure and exhibit a complex temporal behaviour, but
remain ``stable'' being uniformly bounded in time. In such examples various
``eddy'' effects, representing the integral influence of the small-scale
dynamics on the large-scale structures, on the one hand, should play
an important r\^ole in the evolution of the CHM system; on the other, they must
not be introduced ad hoc empirically, but rather follow from the basic
principles. Saturation is impossible if eddy diffusivity is negative. Thus,
although we use the same mathematical approach, our aim is in a sense opposite
to that of Dubrulle and Frisch (1991), Zheligovsky et al. (2001) and
Zheligovsky (2003), who investigated the effects of negative eddy viscosity
and diffusivity as the mechanisms of hydrodynamic and magnetohydrodynamic
instability. The perturbations can only have a complex spatial structure and/or
demonstrate a complex temporal behaviour, if the amplitude equations that are
governing them are themselves complex enough; this is one of the reasons, why
we are obliged to consider a rather general case.

Weakly nonlinear stability to large-scale perturbations of a free convective
hyd\-romagnetic (CHM) regime ${\bf V,H},\cal T$ in a rotating horizontal layer
is considered here. Our immediate goal is to derive amplitude equations
governing the evolution of the perturbations under the assumption that the
$\alpha-$effect is insignificant in the leading order. The amplitude equations
\rf{eq59}, \rf{eq62} and \rf{eqtimedep} (the latter equation emerges if the
perturbed CHM regime is non-steady), that we derive in this paper, involve
a linear {\it operator of combined eddy diffusivity correction} and a nonlinear
{\it operator of anisotropic combined eddy advection}. Both operators are
anisotropic. The system has new qualitative features, compared to the system of
mean-field equations for large-scale perturbations of forced CHM regimes
derived in Zh08. It does not involve mean flow perturbations, and it is mixed:
whereas equations \rf{eq62} for the mean magnetic perturbation are evolutionary,
the remaining ones, \rf{eq59} and \rf{eqtimedep}, do not involve time
derivatives or molecular diffusivity operators. The derivations follow the way
paved in Zh08; consequently, the two papers are similar in structure.

\mi{\bf 2. Equations of hydromagnetic thermal convection,

boundary conditions and the linearisation operators}

\mi
In the Boussinesq approximation the evolution of the perturbation satisfies
the vorticity
\begB
{\partial\bOm\over\partial t}=\nu\nabla^2\bOm+\nabla\times({\bf V}\times\bOm
-{\bf H}\times(\nabla\times{\bf H}))+\nabla\times({\bf V\times\tau e}_3
+\beta{\cal T}{\bf e}_3),
\endB{eq1p5}{CHMB}

\noindent
magnetic induction
\begI
{\partial{\bf H}\over\partial t}=
\eta\nabla^2{\bf H}+\nabla\times({\bf V}\times{\bf H}),
\endI{eq1p2}

\noindent
and heat transfer
$${\partial{\cal T}\over\partial t}=\kappa\nabla^2{\cal T}
-({\bf V}\cdot\nabla){\cal T}$$
equations, as well as the solenoidality conditions
\begI
\nabla\cdot{\bf V}=0,
\endI{eq1p3}\begI
\nabla\cdot{\bf H}=0.
\endI{eq1p8}
(Following Zh08 we use the vorticity equation in place of the momentum equation
to eliminate pressure.) The following notation is used: ${\bf V(x},t)$ and
\begI
\bOm=\nabla\times{\bf V}
\endI{eq1p4}
are velocity and vorticity, respectively, of a flow of an electrically conducting
fluid, ${\bf H}({\bf x},t)$ magnetic field, ${\cal T}({\bf x},t)$ temperature,
$t$ time, $\nu$, $\eta$ and $\kappa$ kinematic, magnetic and thermal molecular
diffusivities, respectively, $\tau/2$ angular velocity of the rotation,
$\beta{\cal T}{\bf e}_3$ the buoyancy force,
${\bf e}_k$ is the unit vector along the axis $x_k$
of the Cartesian coordinate system corotating with the layer.

The CHM regime is {\it free} in the sense that no source terms are present
in the equations (\theCHMB); consequently, the regime is {\it spatially and
temporally invariant}, i.e. there is no dependence of the coefficients in these
equations neither on the position in space, nor on time.

The so called small-angle instability problem (see Podvi\-gina 2008a, 2009
and references therein) is a well-known example of a problem of stability
to large-scale perturbations, in which linear stability of amagnetic convective
regimes (rolls) near the onset is considered. Note, however, that it does not
fall into the class of problems considered here (even if the difference between
linear and weakly nonlinear kinds of stability is neglected), since it involves
two different large scales; like in Zh08, we consider only one large scale.

The following conditions are assumed on the horizontal boundaries of the layer:
\\$\bullet$ no stress:
\begB
\left.{\partial V_1\over\partial x_3}\right|_{x_3=\pm L/2}=
\left.{\partial V_2\over\partial x_3}\right|_{x_3=\pm L/2}=0,\quad
\left.\phantom{|_|}V_3\right|_{x_3=\pm L/2}=0
\endB{eq2SF}{BCs}

\begI
\Rightarrow\left.\phantom{|_|}\Omega_1\right|_{x_3=\pm L/2}
=\left.\phantom{|_|}\!\!\!\Omega_2\right|_{x_3=\pm L/2}=0,\quad
\left.{\partial\Omega_3\over\partial x_3}\right|_{x_3=\pm L/2}=0;
\endI{eq2SFOm}

\noindent
$\bullet$ perfectly electrically conducting boundaries:
\begI
\left.{\partial H_1\over\partial x_3}\right|_{x_3=\pm L/2}=
\left.{\partial H_2\over\partial x_3}\right|_{x_3=\pm L/2}=0,\quad
\left.\phantom{|_|}H_3\right|_{x_3=\pm L/2}=0;
\endI{eq2EC}

\noindent
$\bullet$ isothermal boundaries:
\begI
\left.\phantom{|_|}{\cal T}\right|_{x_3=-L/2}={\cal T}_1,\quad
\left.\phantom{|_|}{\cal T}\right|_{x_3=L/2}={\cal T}_2.
\endL{eq2FT}

\noindent
(Cartesian vector components are enumerated by the subscript.)

It is convenient to introduce a new variable
$\Theta={\cal T}-{\cal T}_1+\delta(x_3+L/2)$, which satisfies the equation
\begin{equation}
{\partial\Theta\over\partial t}=\kappa\nabla^2\Theta
-({\bf V}\cdot\nabla)\Theta+\delta V_3
\label{eq3}\end{equation}
and homogeneous boundary conditions:
\begin{equation}
\left.\phantom{|_|}\Theta\right|_{x_3=\pm L/2}=0.
\label{eq4T}\end{equation}

We define the spatial mean and the oscillatory part of a scalar
or vector field $f$:
\begB
\la f\ra\equiv\lim_{\ell\to\infty}{1\over L\ell^2}\int_{-L/2}^{L/2}
\int_{-\ell/2}^{\ell/2}\int_{-\ell/2}^{\ell/2}f({\bf x})\,{\rm d}x_1\,{\rm d}x_2\,{\rm d}x_3,
\qquad\lb f\rb\equiv f-\la f\ra,
\endB{smean}{means}

\noindent
and the spatio-temporal mean and oscillatory parts of $f$ are

\begI
\lad f\rad\equiv\lim_{\hat{t}\to\infty}
{1\over\hat{t}}\int_0^{\hat{t}}\la f({\bf x},t)\ra\,{\rm d}t,\qquad
\lbd f\rbd\equiv f-\lad f\rad;
\endL{stmean}

\noindent
$\la f\ra_k$ and $\lad f\rad_k$ denote the $k-$th Cartesian
components of the respective means,
\pagebreak
$$\la{\bf f}\ra_v\equiv\la f\ra_3\,{\bf e}_3,\qquad
\lb{\bf f}\rb_v\equiv{\bf f}-\la{\bf f}\ra_v;$$
$$\lad{\bf f}\rad_v\equiv\lad f\rad_3\,{\bf e}_3,\qquad
\lbd{\bf f}\rbd_v\equiv{\bf f}-\lad{\bf f}\rad_v;$$
$$\la{\bf f}\ra_h\equiv\la f\ra_1{\bf e}_1+\la f\ra_2\,{\bf e}_2,\qquad
\lb{\bf f}\rb_h\equiv{\bf f}-\la{\bf f}\ra_h;$$
$$\lad{\bf f}\rad_h\equiv\lad f\rad_1{\bf e}_1+\lad f\rad_2\,{\bf e}_2,\qquad
\lbd{\bf f}\rbd_h\equiv{\bf f}-\lad{\bf f}\rad_h.$$
Following Zh08, the subscripts $v$ and $h$ are thus used to denote vertical and
horizontal components of three-dimensional mean vector fields. The superscripts
$v$ and $h$ have a different meaning: they denote the flow velocity and
magnetic field components of 10-dimensional vector fields
$(\bom,{\bf v,h},\theta)$, and 7-dimensional $(\bom,{\bf h},\theta)$.

Linearisation of equations \rf{eq1p5}, \rf{eq1p2} and \rf{eq3}
in the vicinity of the CHM regime ${\bf V,H},\Theta$ gives rise to the operators
$${\cal L}^\omega(\bom,{\bf v,h},\theta)\equiv-{\partial\bom\over\partial t}
+\nu\nabla^2\bom+\nabla\times\lpar\bf V\times\bom+v\times\bOm$$
\begB
{\bf-H\times(\nabla\times h)-h\times(\nabla\times H})\rpar
+\tau{\partial{\bf v}\over\partial x_3}+\beta\nabla\theta\times{\bf e}_3,
\endB{eq6p11}{defL}\begI
{\cal L}^h({\bf v,h})\equiv-{\partial{\bf h}\over\partial t}+\eta\nabla^2{\bf h}
+\nabla\times({\bf v\times H+V\times h}),
\endI{eq6p2}\begI
{\cal L}^\theta({\bf v},\theta)\equiv-{\partial\theta\over\partial t}
+\kappa\nabla^2\theta-({\bf V}\cdot\nabla)\theta-({\bf v}\cdot\nabla)\Theta+\delta v_3.
\endL{eq6p3}

\noindent
The operator of linearisation of the system (\theCHMB) is
${\cal L}=({\cal L}^\omega,{\cal L}^h,{\cal L}^\theta)$.

We consider weakly nonlinear regimes of perturbations, the amplitude of which
is of the order of $\varepsilon$. The perturbed state $\bOm+\varepsilon\bom$,
$\bf V+\varepsilon v$, $\bf H+\varepsilon h$, $\Theta+\varepsilon\theta$
also satisfies (\theCHMB) and \rf{eq3}, and hence the profiles of perturbations
$\bom,{\bf v,h},\theta,p$ (henceforth called perturbations) satisfy the equations
\begB
{\cal L}^\omega(\bom,{\bf v,h},\theta)
+\varepsilon\nabla\times({\bf v}\times\bom-{\bf h}\times(\nabla\times{\bf h}))={\bf 0},
\endB{eq5om}{EQpert}\begI
{\cal L}^h({\bf v,h})+\varepsilon\nabla\times({\bf v\times h})={\bf 0},
\endI{eq5p2}\begI
{\cal L}^\theta({\bf v},\theta)-\varepsilon({\bf v}\cdot\nabla)\theta=0,
\endI{eq5p3}\begI
\nabla\cdot{\bf h}=0,
\endI{eq5p4}\begI
\nabla\cdot\bom=\nabla\cdot{\bf v}=0,
\endI{eq5p5}\begI
\nabla\times{\bf v}=\bom.
\endI{eq5p6}
Averaging of the horizontal component of the Navier-Stokes equation over
the fluid volume yields
$${\partial\la{\bf v}\ra_h\over\partial t}=
\la{\bf v}\ra_h\times\tau{\bf e}_3-\la\nabla p\ra_h.$$
We assume that there is no mean horizontal fluid flow through the layer,
i.e. at any time $\la\nabla p\ra_h=\bf 0$ and
\begI
\la{\bf v}\ra_h={\bf 0}.
\endL{eq5p7}

\mi{\bf 3. Large-scale weakly nonlinear perturbations of CHM regimes}

\mi
We introduce the {\it slow} spatial, ${\bf X}=\varepsilon(x_1,x_2)$, and
temporal, $T=\varepsilon^2t$, variables. (We assume that the layer width $L$
is of the order of unity, and hence a slow variable in the vertical direction
is not introduced.) For vector fields depending both on
fast and slow variables, integration in the definitions
(\themeans) of mean fields is performed only over the small-scale fast
variables $\bf x$ or $({\bf x},t)$. The exponent in the temporal scale ratio
is suitable for CHM regimes with an insignificant $\alpha-$effect.
A power series solution to the problem (\theEQpert) is sought:
\begin{equation}\begin{array}{c}
\displaystyle\bom=\sum_{n=0}^\infty\bom_n({\bf x},t,{\bf X},T)\,\varepsilon^n,\qquad
{\bf v}=\sum_{n=0}^\infty{\bf v}_n({\bf x},t,{\bf X},T)\,\varepsilon^n,\\
\displaystyle{\bf h}=\sum_{n=0}^\infty{\bf h}_n({\bf x},t,{\bf X},T)\,\varepsilon^n,\qquad
\theta=\sum_{n=0}^\infty\theta_n({\bf x},t,{\bf X},T)\,\varepsilon^n.
\label{pertser}\end{array}\end{equation}
Solenoidality conditions for the flow, magnetic field and vorticity
result in the following equations: for any $n\ge0$,
\begB
\nabla_{\bf X}\cdot\la{\bf v}_n\ra_h=\nabla_{\bf X}\cdot\la{\bf h}_n\ra_h=0,
\endB{eq17p1}{termsolA}\begI
\nabla_{\bf x}\cdot\lb{\bf v}_n\rb_h+\nabla_{\bf X}\cdot\lb{\bf v}_{n-1}\rb_h=0,
\endI{eq17p2}\begI
\nabla_{\bf x}\cdot\lb{\bf h}_n\rb_h+\nabla_{\bf X}\cdot\lb{\bf h}_{n-1}\rb_h=0,
\endI{eq17p3}\begI
\nabla_{\bf x}\cdot\lb\bom_n\rb_v+\nabla_{\bf X}\cdot\lb\bom_{n-1}\rb_v=0.
\endL{eq17p5}
(In differential operators with the indices $\bf x$ and $\bf X$
differentiation in the respective spatial fast and slow variables is performed;
$\nabla_{\bf X}=(\partial/\partial X_1,\partial/\partial X_2,0)$.
Differentiation in fast variables only is assumed in the definitions (\thedefL)
of linearisation operators. All quantities for $n<0$ are zero by definition.)

The identity \rf{eq5p6} translates into the equation
\begin{equation}
\nabla_{\bf x}\times{\bf v}_n=\bom_n-\nabla_{\bf X}\times{\bf v}_{n-1}.
\label{eqomega}\end{equation}
Equations \rf{eq17p2} and \rf{eq17p3}, together with \rf{eqomega} where the
index is $n$ changed to $n-1$, implies that the right-hand side of \rf{eqomega}
is solenoidal in fast variables. The solvability condition for \rf{eqomega} is
\begin{equation}
\la\bom_n\ra_v=\nabla_{\bf X}\times\la{\bf v}_{n-1}\ra_h.
\label{eq19}\end{equation}
This equation has a solution, if the mean of $\la\bom_n\ra_v$ over
the plane of slow variables vanishes. Equation \rf{eqomega} then reduces to
\begin{equation}
\nabla_{\bf x}\times\lb{\bf v}_n\rb_h=\lb\bom_n\rb_v
-\nabla_{\bf X}\times\lb{\bf v}_{n-1}\rb_h,
\label{eq30r}\end{equation}
and a solution $\lb{\bf v}_n\rb_h$ can be found as a sum
of solutions to two Neumann problems (see Zh08).

A hierarchy of equations (A.1)--(A.3) (derived in Zh08
and presented here in Appendix A for reader's convenience) emerges
at different orders of $\varepsilon$ upon substitution of the series
\rf{pertser} into \xrf{eq5om}{eq5p3} and expansion of the resulting equation
in power series in $\varepsilon$. We will solve them successively, essentially
following the procedure described in Sections 5-9 of Zh08. For $n>0$ the order
of calculations
is not quite straightforward, for each $n$ it involves the following steps:\\
$1^\circ.$ Derive a PDE in slow variables in $\lad\bom_{n-1}\rad_v$ and
find $\la{\bf v}_n\ra_h-\lad{\bf v}_n\rad_h$ using the mean and
oscillatory, respectively, part of the spatial average of the vertical
component of (A.1) at order $\varepsilon^{n+1}$.\\
$2^\circ.$ Derive a PDE in slow variables in $\lad{\bf h}_{n-2}\rad_h$
and calculate $\la{\bf h}_n\ra_h-\lad{\bf h}_n\rad_h$ using the mean and
oscillatory, respectively, part of the spatial average of the horizontal
component of (A.2) at order $\varepsilon^n$.\\
$3^\circ.$ Find the potential parts (in fast variables) of
$\bom_n,{\bf v}_n$ and ${\bf h}_n$, applying (\thetermsolA).\\
$4^\circ.$ Applying results of the previous steps, derive from (A.1)--(A.3)
a system of PDE's in fast variables in solenoidal parts of
$\lb\bom_n\rb_v,\lb{\bf v}_n\rb_h,{\bf h}_n$ and $\theta_n$.\\
$5^\circ.$ Applying two solvability conditions, calculate $\lad{\bf v}_n\rad$.
(The averages considered at steps $1^\circ$ and $2^\circ$ also represent
solvability conditions for the system (A.1)--(A.3), see Section 5.)
Apply the remaining solvability conditions, if any are not yet satisfied.\\
$6^\circ.$ Use the system of PDE's derived at the previous step
to calculate $\bom_n,{\bf v}_n,{\bf h}_n$ and $\theta_n$
in the terms of solutions to auxiliary problems and the yet unknown
spatio-temporal averages of the terms of expansions \rf{pertser}.\\
At this stage solution of the system (A.1)--(A.3) for the considered index
$n$ is completed, and we continue by solving the equations at the next order
$\varepsilon^{n+1}$ starting anew with step $1^\circ.$ Since the equations for
$n=0$ studied in the next section are homogenous, they represent a special
case, in which the flow of operations is slightly altered. In what follows,
we indicate at each point, which step is being carried out. Although we stop
as soon as derive the amplitude equations, further application of the procedure
would yield solutions to an arbitrary number of successive systems (A.1)--(A.3)
in the hierarchy. Thus a complete asymptotic expansion \rf{pertser} can be
constructed.

\mi{\bf 4. Order $\varepsilon^0$ equations}

\mi
Equations (A.1)--(A.3), \rf{eqomega} and solenoidality conditions
\xrf{eq17p2}{eq17p5} reduce for $n=0$ to
\begB
{\cal L}(\bom_0,{\bf v}_0,{\bf h}_0,\theta_0)={\bf 0},
\endB{eq20}{ordernull}\begI
\bom_0=\nabla_{\bf x}\times{\bf v}_0,\qquad
\nabla\cdot\bom_0=\nabla\cdot{\bf v}_0=\nabla\cdot{\bf h}_0=0.
\endL{sol0}
Differentiation of \rf{eq1p5}, \rf{eq1p2} and \rf{eq3} in $x_k,\ k=1,2$, or
in time $t$ demonstrates that ${\bf S}_k^{\cdt,x}\equiv(\partial\bOm/\partial x_k,
\partial{\bf H}/\partial x_k,\partial\Theta/\partial x_k)$ and
${\bf S}^{\cdt,t}\equiv(\partial\bOm/\partial t,\partial{\bf H}/\partial t,
\partial\Theta/\partial t)$ are solutions to (\theordernull). They satisfy the
boundary conditions, whichever of them (no stress or no slip boundaries, built
of perfectly conducting or a dielectric material, isothermal or transferring
a prescribed heat flow) are chosen for the flow, magnetic field
and temperature. This is a consequence of the assumed spatial and temporal
invariance of the equation, governing the CHM regime ${\bf V,H},\Theta$.
The three solutions to (\theordernull) are not guaranteed to be linearly
independent; for instance, ${\bf S}^{\cdt,t}={\bf 0}$ for steady CHM regimes. Also,
${\bf S}_k^{\cdt,x}$ are linearly dependent, if the CHM state is independent
of a horizontal variable $a_1x_1+a_2x_2$ for some constant $a_1$ and $a_2$,
not vanishing together; we will not consider such essentially two-dimensional
CHM regimes.

Whether other solutions to (\theordernull) exist depends on the assumed
boundary conditions and on the parameter values. In what follows we consider
boundary conditions \rf{eq2SF}, \rf{eq2EC} and \rf{eq2FT}. They are inherently
consistent with the solenoidality conditions and thereby they are widely used
in numerical studies of thermal convection and convective dynamos (see, e.g.,
Podvigina 2006, 2008b).

Following Zh08, we introduce the operator of linearisation in the form
$${\cal M}(\bom,{\bf h},\theta)\equiv{\cal L}(\bom,{\cal R}\bom,{\bf h},\theta).$$
Here $\cal R$ is the inverse curl, acting from the space of
solenoidal vector fields globally bounded in the layer and satisfying
\rf{eq2SFOm}, with a zero spatial mean of the vertical component,
into the space of solenoidal vector fields satisfying \rf{eq2SF}, with a zero
spatial mean of the horizontal component (see Zh08, Section 3). The domain of
$\cal M$, which we denote by $\cal D$, consists of vector fields
$(\bom,{\bf h},\theta)$, globally bounded in the
layer, satisfying the respective boundary conditions, and such that
$\bom$ and $\bf h$ are solenoidal with $\la\bom\ra_v=\bf 0$.

The operator adjoint to $\cal M$ is
${\cal M}^*=(({\cal M}^*)^\omega,({\cal M}^*)^h,({\cal M}^*)^\theta)$,
$$({\cal M}^*)^\omega(\bom,{\bf h},\theta)={\partial\bom\over\partial t}
+\nu\nabla^2\bom+{\cal R}'\lb-\nabla\times({\bf V}\times(\nabla\times\bom))$$
$$+{\bf H}\times(\nabla\times{\bf h})+\bOm\times(\nabla\times\bom)
-\tau{\partial\bom\over\partial x_3}+\delta\theta{\bf e}_3-\theta\nabla\Theta\rb_h,$$
$$({\cal M}^*)^h(\bom,{\bf h})={\partial{\bf h}\over\partial t}+\eta\nabla^2{\bf h}
+\nabla\times({\bf H}\times(\nabla\times\bom))+{\cal P}((\nabla\times\bom)\times(\nabla\times{\bf H})
-{\bf V}\times(\nabla\times{\bf h})),$$
$$({\cal M}^*)^\theta(\bom,{\bf h},\theta)={\partial\theta\over\partial t}
+\kappa\nabla^2\theta+({\bf V}\cdot\nabla)\theta+\beta{\bf e}_3\cdot(\nabla\times\bom).$$
Here the operator ${\cal R}'$ is another ``inverse curl'':
${\cal R}'{\bf v}=\bom$ is a solution to the problem
$$\nabla^2\bom=-\nabla\times{\bf v},\qquad\nabla\cdot\bom=0,\qquad\la\bom\ra_v=\bf 0$$
satisfying \rf{eq2SFOm}, and $\cal P$ is the projection of a three-dimensional
vector field into the subspace of solenoidal fields with a zero spatial mean of
the horizontal component, satisfying boundary conditions \rf{eq2EC} (see
Section 3 in Zh08 for a detailed explanation).
The operators $\cal M$ and ${\cal M}^*$ turn out to have the same domain.

Clearly, ${\cal M}^*({\bf 0},{\bf C}^h,0)=\bf 0$, where ${\bf C}^h$ is
a constant horizontal vector. Consequently, there exist eigenmodes
${\bf S}^\cdt=({\bf S}^\omega,{\bf S}^h,S^\theta)$ of the operator
$\cal M$ associated with the zero eigenvalue, for which
$\lad{\bf S}^h\rad_h\ne\bf 0$. Since spatio-temporal means of ${\bf S}_k^{\cdt,x}$
and ${\bf S}^{\cdt,t}$ are zero, the latter neutral modes are distinct. We denote by
${\bf S}^{\cdt,h}_k=({\bf S}^{\omega,h}_k,{\bf S}^{h,h}_k,S^{\theta,h}_k)$
the neutral modes with non-zero horizontal magnetic components, normalised so
that $\lad{\bf S}^{h,h}_k\rad_h\!={\bf e}_k$.

\underline{Step $2^\circ$ for $n=0$.} Averaging in space the
horizontal magnetic component of \rf{eq20} demonstrates that the spatial
means of the horizontal magnetic field components are time-independent; thus
$\la{\bf h}_0\ra_h=\lad{\bf h}_0\rad_h$.
The normalisation is equivalent to the conditions
$\la{\bf S}^{h,h}_k\ra_h={\bf e}_k$ at any time $t\ge 0$. Spatial means of
horizontal components of the flows ${\bf S}^{v,h}_k={\cal R}{\bf S}^{\omega,h}_k$
are zero.

Thus, we have identified up to 5 different solutions to (\theordernull) in $\cal D$
existing for any parameter values: ${\bf S}_k^\cdt={\bf S}_k^{\cdt,x}\ (k=1,2)$,
${\bf S}_{k+2}^\cdt={\bf S}_k^{\cdt,h}$, and if the CHM regime ${\bf V,H},\Theta$
is unsteady ${\bf S}_5^\cdt={\bf S}^{\cdt,t}$. Consequently,
the kernel of ${\cal M}^*$ is spanned by 4 or 5 eigenfunctions
${\bf S}^{*\cdt}_k$: ${\bf S}^{*\cdt}_{k+2}=({\bf 0},{\bf e}_k,0)$ for $k=1,2$,
and ${\bf S}^{*\cdt}_5$ exists for non-steady CHM regimes. ${\bf S}^{*\cdt}_k$
can be chosen to be biorthogonal to ${\bf S}^\cdt_j$ and normalised so that
\begin{equation}
\lad{\bf S}^\cdt_k\cdot{\bf S}^{*\cdt}_j\rad=\delta^k_j
\label{norm125}\end{equation}
for $1\le k,j\le K$. Here $K$ is the dimension of $\ker\cal M$,
$\delta^k_j$ is the Kronecker symbol, and the scalar product of
7-dimensional vector fields $({\bf S}^\omega,{\bf S}^h,S^\theta)$ is assumed.
The means $\la{\bf S}^{*v}_k\ra_h$ vanish
for all $k\le 5$, $\lad{\bf S}^{*h}_k\rad_h=\bf 0$ for $k=1,2,5$.

\underline{Step $6^\circ$ for $n=0$.}
By linearity a solution to (\theordernull) can be expressed as
\begin{equation}
(\bom_0,{\bf h}_0,\theta_0)=\bxi_0^\cdt+\sum_{k=1}^Kc_{0,k}{\bf S}^\cdt_k
\label{eq23}\end{equation}
(interpreted as an equality of 7-dimensional vectors). Here
$\bxi^\cdt_0({\bf x,X},t,T)=(\bxi_0^v,\bxi_0^h,\xi_0^\theta)$ is a transient
also satisfying the equations (\theordernull): initial
conditions for $\bxi^\cdt_0$, which can be found from \rf{eq23} at $t=0$,
must belong to the stable manifold of the perturbed CHM state
${\bf V,H},\Theta$, i.e. $\bxi^\cdt_0$ must exponentially decay in time $t$.
Our task is to construct a closed system of equations for the leading order
amplitudes $c_{0,k}({\bf X},T)$.

If the CHM regime ${\bf V,H},\Theta$ possesses a symmetry about the vertical
\vspace*{1pt}
axis or parity invariance, perhaps involving a time shift $\widetilde T\ge0$
(see the definitions in Section 7 in Zh08), then symmetric and antisymmetric sets
of fields are invariant subspaces for the linearisation operator $\cal M$.
Thus any eigenmode is a symmetric or antisymmetric set
of vector fields (if exponentially decaying transients of arbitrary symmetry
are ignored in the case of time-dependent modes). In particular,
all ${\bf S}^\cdt_k$ are antisymmetric sets, except for ${\bf S}^{\cdt,t}$,
which is symmetric. Consequently, one can choose a basis in $\ker{\cal M}^*$,
where all vector fields are antisymmetric sets, except for a symmetric set
${\bf S}^{*\cdt}_K$, if the CHM state ${\bf V,H},\Theta$ is unsteady.

\mi{\bf 5. Solvability conditions}

\mi
In this section we consider the system of equations
\begB
{\cal M}(\bom,{\bf h},\theta)=({\bf f}^\omega,{\bf f}^h,f^\theta),\quad
\nabla_{\bf x}\cdot\bom=\nabla_{\bf x}\cdot{\bf h}=
\nabla_{\bf x}\cdot{\bf f}^\omega=\nabla_{\bf x}\cdot{\bf f}^h=0,
\endB{MEQ1}{Msys}\begI
\la\bom\ra_v={\bf 0}.
\endL{MEQ5}
A solution to (\theMsys) exists only, if the following solvability conditions
are satisfied:
\begin{equation}
\la{\bf f}^\omega\ra_v={\bf 0}
\label{solveomega}\end{equation}
(stemming from \rf{MEQ5}~),
\begin{equation}
\lad{\bf f}^h\rad_h={\bf 0}
\label{solveh}\end{equation}
and
\begin{equation}
\lad{\bf f}\cdot{\bf S}^{*\cdt}_k\rad=0
\label{solves}\end{equation}
for $k=1,2,5$. Equation \rf{solveh} is identical to relations \rf{solves}
for $k=3,4$. (If the CHM state ${\bf V,H},\Theta$ is steady, then \rf{solves}
is not considered for $k=5$.)

We assume henceforth that the invariant subspace spanned by
${\bf S}^\cdt_k$, $1\le k\le K$, is the complete kernel of $\cal M$,
where $K$ is the number of eigenmodes in the basis in $\ker\cal M$: $K=4$ or 5
for the steady and non-steady CHM states ${\bf V,H},\Theta$, respectively.
In the absence of space periodicity the conditions \xrf{solveomega}{solves}
may be insufficient for existence of a globally bounded solution to (\theMsys):
already quasi-periodicity in horizontal directions can be problematic.
Nevertheless we do not impose periodicity conditions, which is too restrictive,
but assume that the auxiliary problems of the form (\theMsys) do have
solutions in $\cal D$, if \xrf{solveomega}{solves} hold true. Subtracting
an appropriate linear combination of ${\bf S}^\cdt_k,\ k=3,4$, which belong to
$\ker\cal M$, one obtains a solution to (\theMsys) with $\lad{\bf h}\rad_h=\bf 0$.

For $n=0$, \rf{eq19} reduces to $\la\bom_0\ra_v=\bf 0$, which
is compatible with (\theordernull) by virtue of the identity
\begin{equation}
\la{\cal L}^\omega(\bom,{\bf v,h},\theta)\ra_v=-{\partial\la\bom\ra_v\over\partial t}.
\label{eq11p1}\end{equation}
The solvability conditions can be used to demonstrate that, unlike in the case
of forced convection, $\la{\bf v}_0\ra_h=\bf 0$.

\underline{Step $1^\circ$ for $n=0$.} From (A.4) for $n=1$,
$${\partial\la\bom_1\ra_v\over\partial t}={\bf 0},\quad\Rightarrow\quad
\la\bom_1\ra_v=\lad\bom_1\rad_v,$$
and hence \rf{eq19} for $n=1$ together with \rf{eq17p1} for $n=0$ implies
$$\la{\bf v}_0\ra_h=\lad{\bf v}_0\rad_h+{\bf v}'_0(t,T).$$
However, $\la\,\lb{\bf v}_0\rb_h|_{{\bf X}=\varepsilon(x_1,x_2)}\ra$ is
asymptotically smaller than any power of $\varepsilon$ (see Zh08,
Appendix C). Thus \rf{eq5p7} requires ${\bf v}'_0=\bf 0$, i.e.
\begin{equation}
\la{\bf v}_0\ra_h=\lad{\bf v}_0\rad_h,
\label{spmean}\end{equation}
and vanishing of the average of $\lad{\bf v}_0\rad_h$ over slow
spatial variables.

\underline{Step $3^\circ$ for $n=0$.} Consequently, \rf{eq20} reduces to
\begin{equation}
{\cal M}(\bom_0,{\bf h}_0,\theta_0)=
(\lad{\bf v}_0\rad_h\cdot\nabla_{\bf x})(\bOm,{\bf H},\Theta).
\label{Morder1}\end{equation}

\underline{Step $4^\circ$ for $n=0$.}
Scalar multiplying this equation by ${\bf S}^{*\cdt}_j$ and averaging
the result, we find that
$$\ladb{\bf S}^{*\cdt}_j\cdot\lpar(\lad{\bf v}_0\rad_h\cdot\nabla_{\bf x})
(\bOm,{\bf H},\Theta)\rpar\radb=0$$
is a necessary condition for existence of its solution.
Hence, biorthogonality \rf{norm125} of ${\bf S}^\cdt_k$ to ${\bf S}^{*\cdt}_j$
for $k,j=1,2$ implies, together with \rf{spmean} and \rf{eq19} for $n=1$,
\begin{equation}
\la{\bf v}_0\ra_h=\la\bom_1\ra_v={\bf 0}.
\label{avom1}\end{equation}
In other words, solvability of \rf{Morder1} in $\cal D$ requires that the
mean perturbation flow vanishes. We further discuss this issue in Section 10.

\mi{\bf 6. Order $\varepsilon^1$ and $\varepsilon^2$ equations:
$\alpha-$effect in the leading order}

\mi
In this section we derive the $\alpha-$effect operators, present in the CHM
system under consideration.

\underline{Step $1^\circ$ for $n=1$.} Substitution of the flow and magnetic
field \rf{eq23} into (A.4) for $n=2$ yields
\begin{equation}
{\partial\la\bom_2\ra_v\over\partial t}=\nabla_{\bf X}\times
\left(\,\sum_{k=1}^K\sum_{m=1}^2\bal^\omega_{m,k}
{\partial c_{0,k}\over\partial X_m}+\txi{\omega}\right),
\label{eq30}\end{equation}
where
\begin{equation}
\bal^\omega_{m,k}=\la{\bf S}^h_kH_m+{\bf H}(S^h_k)_m-{\bf S}^v_kV_m-{\bf V}(S^v_k)_m\ra_h,
\label{alome}\end{equation}
$$\txi{\omega}=\la{\bf V}\times(\nabla_{\bf X}\times\bxi^v_0)
-{\bf V}\nabla_{\bf X}\cdot\bxi^v_0-{\bf H}\times(\nabla_{\bf X}\times\bxi^h_0)
+{\bf H}\nabla_{\bf X}\cdot\bxi^h_0\ra_h.$$
The differential operator in the right-hand side of \rf{eq30} represents
the AKA--effect (anisotropic kinematic $\alpha-$effect) operator
in the leading order. Integrating \rf{eq30} in fast time, find
\begin{equation}
\la\bom_2\ra_v=\lad\bom_2\rad_v+\nabla_{\bf X}\times\left(\,
\sum_{k=1}^K\sum_{m=1}^2\lbdb\int_0^t\bal^\omega_{m,k}\,{\rm d}t\rbdb
{\partial c_{0,k}\over\partial X_m}+\lbdb\int_0^t\txi{\omega}{\rm d}t\rbdb\right).
\label{eq39}\end{equation}
Thus $\la\bom_2\ra_v$ is well-defined, if the means
$\lad\int_0^t\bal^\omega_{m,k}\,{\rm d}t\rad$ are. This is {\it the condition of
insignificance of the AKA--effect} in the leading order. It is stronger than
the solvability condition $\lad{\bf f}^\omega\rad_v=\bf 0$ for the equation
${\cal L}^\omega(\bom,{\bf v,h},\theta)={\bf f}^\omega$, which follows
from \rf{eq11p1} and in the case of (A.1) for $n=2$ is
\begin{equation}
\lad\bal^\omega_{m,k}\rad={\bf 0}.
\label{eq34}\end{equation}

The solution to equations \rf{eq19} for $n=2$ and \rf{eq17p1} for $n=1$ is
therefore
\begin{equation}
\la{\bf v}_1\ra_h\!=\txi{v}\!+\lad{\bf v}_1\rad_h\!+\!\sum_{k=1}^K\sum_{m=1}^2\sum_{j=1}^2
\lbdb\int_0^t{(\alpha^\omega_{m,k})}_{\!j}\,{\rm d}t\rbdb\left(
{\partial c_{0,k}\over\partial X_m}\,{\bf e}_j-\nabla_{\bf X}
{\partial^2\nabla^{-2}_{\bf X}c_{0,k}\over\partial X_m\partial X_j}\right),
\label{avevI}\end{equation}
where $\nabla^{-2}_{\bf X}$ is the inverse Laplacian in slow variables,
acting from the space of globally bounded functions, whose average
in $\bf X$ is zero, into the same space;
$$\txi{v}=\lbdb\int_0^t\bxi'{\rm d}t\rbdb,\qquad
\bxi'({\bf X},t,T)=\txi{\omega}-\nabla_{\bf X}\nabla^{-2}_{\bf X}(\nabla_{\bf X}\cdot\txi{\omega}).$$

\underline{Step $2^\circ$ for $n=1$.}
A possible presence of magnetic $\alpha-$effect in the CHM system with
electrically conducting boundaries \rf{eq2EC} can be revealed similarly.
Averaging of the horizontal component of (A.2) for $n=1$
in fast spatial variables with the use of the identity
\begin{equation}
\la{\cal L}^h({\bf v,h})\ra_h=-{\partial\la{\bf h}\ra_h\over\partial t}
\label{eq11p2}\end{equation}
and substitution of the flow and magnetic component of \rf{eq23} yields
\begin{equation}
{\partial\la{\bf h}_1\ra_h\over\partial t}=\nabla_{\bf X}\times
\left(\,\sum_{k=1}^K\bal^h_kc_{0,k}+\txi{h}\right),
\label{eq35}\end{equation}
where
\begin{equation}
\bal^h_k=\la{\bf V}\times{\bf S}^h_k+{\bf S}^v_k\times{\bf H}\ra_v,
\label{alh}\end{equation}
$$\txi{h}=\la\bxi^v_0\times{\bf H}+{\bf V}\times\bxi^h_0\ra_v.$$
The differential operator in the right-hand side of \rf{eq35} represents
the magnetic $\alpha-$effect. The equation yields
\begin{equation}
\la{\bf h}_1\ra_h=\lad{\bf h}_1\rad_h+\nabla_{\bf X}\times\left(\,
\sum_{k=1}^K\lbdb\int_0^t\bal^h_k\,{\rm d}t\rbdb c_{0,k}
+\lbdb\int_0^t\!\txi{h}{\rm d}t\rbdb\right)\!.
\label{Malpha}\end{equation}
{\it Magnetic $\alpha-$effect is insignificant} in the leading order, if
the means $\lad\int_0^t\bal^h_k{\rm d}t\rad$ exist and as a result
$\la{\bf h}_1\ra_h$ is well-defined by \rf{Malpha}. The condition of
insignificance of the magnetic $\alpha-$effect is stronger than the solvability
condition $\lad{\bf f}^h\rad_h=\bf 0$ for the equation
${\cal L}^h({\bf v,h},\theta)={\bf f}^h$, which follows from
\rf{eq11p2} and in the case of (A.2) for $n=1$ is
\begin{equation}
\lad\bal^h_k\rad={\bf 0}.
\label{eq37}\end{equation}

If the perturbed CHM state is steady or periodic in time, the $4K$ scalar
relations \rf{eq34} are sufficient for insignificance of kinematic
$\alpha-$effect, and the $K$ scalar relations \rf{eq37} are sufficient for
insignificance of the magnetic $\alpha-$effect.
Elementary algebra shows, that these conditions hold true for $k=1,2,5$:
$$\bal^\omega_{m,k}=\bal^h_k={\bf 0}\hbox{ for }k=1,2,\qquad
\bal^\omega_{m,5}={\partial\over\partial t}\,\la{\bf H}H_m-{\bf V}V_m\ra_h,\qquad
\bal^h_5={\partial\over\partial t}\,\la{\bf V}\times{\bf H}\ra_v.$$

If the CHM regime ${\bf V,H},\Theta$ has a symmetry considered in Zh08, then
\rf{eq34} and \rf{eq37} are satisfied for all $k$, and thus the AKA-- and magnetic
$\alpha-$effects are insignificant in the leading order. If the symmetry is
spatial (i.e. the time shift $\widetilde T$ is zero), then $\bal^\cdt_{\cdtx}=\bf 0$
for all $k\le K=4$; if it is spatio-temporal, then for all $k\le 4$
$\bal^\cdt_{\cdtx}(t+\widetilde T)=-\bal^\cdt_{\cdtx}(t)$
and $\lbd\int_0^t\bal^\cdt_{\cdtx}\,{\rm d}t\rbd$
satisfies the same relation (see the proof in Zh08, Section 7); the remaining
vector fields $\bal^\cdt_{\cdtx}$ for $k=5$ satisfy
$\bal^\cdt_{\cdtx}(t+\widetilde T)=\bal^\cdt_{\cdtx}(t)$, and hence
$\lbd\int_0^t\bal^\cdt_{\cdtx}\,{\rm d}t\rbd$ also satisfy the latter relation.

The terms $\lbd\int_0^t\txi{\omega}{\rm d}t\rbd$ in \rf{eq39} and
$\lbd\int_0^t\txi{h}{\rm d}t\rbd$ in \rf{Malpha} are not problematic, because
$\txi{\omega}$ and $\txi{h}$ exponentially decay and thus the means
$\lad\int_0^t\txi{\omega}{\rm d}t\rad$ and $\lad\int_0^t\txi{h}{\rm d}t\rad$ are
well-defined. The quantities $\lbd\int_0^t\txi{\omega}{\rm d}t\rbd$,
$\txi{v}$ and $\lbd\int_0^t\txi{h}{\rm d}t\rbd$ also decay exponentially
(see Zh08, Appendix D).

The multiscale approach remains feasible even, if \rf{eq34} and/or \rf{eq37} do
not hold true. It is well-known (see Dubrulle and Frisch 1991), that in this case
another slow time scale is appropriate, $T=\varepsilon t$. Then the new term
$\partial\la{\bf h}_0\ra_h/\partial T$ emerging in the left-hand side of
\rf{eq35} balances the magnetic $\alpha-$effect term. Then equations \rf{eq30}
and \rf{eq35}, supplemented by \rf{eqtimedep} (see the next section) if the CHM
regime ${\bf V,H},\Theta$ is unsteady, constitute a closed system of linear
first-order PDE's (together with the solenoidality conditions \rf{eq17p1}, for
$n=0$ for the mean magnetic field, and for $n=1$ for the flow \rf{avev1} also
derived in the next section; in view of \rf{avom1}, \rf{eqtimedep} and
\rf{avev1} are not affected by the change in the time scaling). This
system turns out to be mixed: while \rf{eq35} is an evolutionary equation,
\rf{eq30} and \rf{eqtimedep} are not. Solutions to such systems generically
exhibit unbounded exponential growth. (Multiscale expansion can yield amplitude
equations, where the $\alpha-$effect coexists together with the molecular
diffusivity and the nonlinear advection, but this requires construction of the
asymptotic expansion along a different line in the parameter space. For
instance, Frisch et al., 1987, obtained such mean-flow equations for
perturbations of flows in a hydrodynamic weakly nonlinear stability problem
in the limit of small Reynolds numbers $R$, assuming
that the scale ratios were linked to $R$.) Consequently, from now on we
focus on a potentially more interesting case of insignificant $\alpha-$effect,
where possible growth of perturbations may saturate due to nonlinearity.

\mi{\bf 7. Order $\varepsilon^1$ equations: $\alpha-$effect, insignificant in the leading order}

\mi
Assuming that the AKA-- and magnetic $\alpha-$effects are insignificant,
in this section we solve equations (A.1)--(A.3) for $n=1$.

\underline{Step $3^\circ$ for $n=1$.} To apply the
solvability conditions \xrf{solveomega}{solves}, this system must be
transformed to take the form of the problem (\theMsys). Thus, the gradient
parts of $\bom_1,{\bf v}_1$ and ${\bf h}_1$, and the mean part of ${\bf v}_1$
must be isolated. These quantities can be found using \rf{eq30r},
\xrf{eq17p2}{eq17p5} for $n=1$, \rf{avom1} and \rf{eq23}:
\begB
\bom_1=\sum_{k=1}^K\sum_{m=1}^2{\partial c_{0,k}\over\partial X_m}
\,{\bf e}_m\times{\bf S}^v_k+\bom'_1,
\endB{omone}{vhomone}

\begI
{\bf v}_1=\la{\bf v}_1\ra_h+\sum_{k=1}^K\sum_{m=1}^2{\partial c_{0,k}\over\partial X_m}
\,\nabla_{\bf x}s^v_{m,k}+{\bf v}'_1,
\endI{vone}

\begI
{\bf h}_1=\sum_{k=1}^K\sum_{m=1}^2{\partial c_{0,k}\over\partial X_m}
\,\nabla_{\bf x}s^h_{m,k}+{\bf h}'_1,
\endL{hone}
where
\begin{equation}
\nabla_{\bf x}\cdot\bom'_1=\nabla_{\bf x}\cdot{\bf v}'_1
=\nabla_{\bf x}\cdot{\bf h}'_1=0,\qquad\la\bom'_1\ra_v=\la{\bf v}'_1\ra_h={\bf 0},\qquad
\nabla_{\bf x}\times{\bf v}'_1=\bom'_1
\label{vprime}\end{equation}
and $s^\cdt_{m,k}({\bf x},t)$ are globally bounded solutions to Neumann problems
$$\nabla^2_{\bf x}\,s^v_{m,k}=-(S^v_k)_m,\quad
\left.{\partial s^v_{m,k}\over\partial x_3}\right|_{x_3=\pm L/2}=0;$$
$$\nabla^2_{\bf x}\,s^h_{m,k}=-{\lb S^h_k\rb}_m,\quad
\left.{\partial s^h_{m,k}\over\partial x_3}\right|_{x_3=\pm L/2}=0.$$
Consequently, $\bom'_1,{\bf v}'_1$ and ${\bf h}'_1$ satisfy the same boundary
conditions, as $\bom_1,{\bf v}_1$ and ${\bf h}_1$, respectively.

\underline{Step $4^\circ$ for $n=1$.}
Upon substitution of (\thevhomone), equations (A.1)--(A.3) for $n=1$ take the form
$${\cal M}^\omega(\bom'_1,{\bf h}'_1,\theta_1)=-\sum_{k=1}^K\sum_{m=1}^2
{\cal L}^\omega({\bf e}_m\times{\bf S}^v_k,\nabla_{\bf x}s^v_{m,k},
\nabla_{\bf x}s^h_{m,k},0)\,{\partial c_{0,k}\over\partial X_m}
+(\la{\bf v}_1\ra_h\cdot\nabla_{\bf x})\bOm$$
$$-\,2\nu(\nabla_{\bf x}\cdot\nabla_{\bf X})\bom_0
-\nabla_{\bf X}\times\left({\bf V}\times\bom_0+{\bf v}_0\times\bOm
-{\bf H\times(\nabla_x\times h}_0)-{\bf h}_0\times(\nabla_{\bf x}\times{\bf H})\right)$$
\begB
-\nabla_{\bf x}\times(-{\bf H}\times(\nabla_{\bf X}\times{\bf h}_0)
+{\bf v}_0\times\bom_0-{\bf h}_0\times(\nabla_{\bf x}\times{\bf h}_0))
-\beta\nabla_{\bf X}\theta_0\times{\bf e}_3,
\endB{MorderNomega}{MorderN}
$${\cal M}^h(\bom'_1,{\bf h}'_1)=-\sum_{k=1}^K\sum_{m=1}^2{\cal L}^h
(\nabla_{\bf x}s^v_{m,k},\nabla_{\bf x}s^h_{m,k})\,{\partial c_{0,k}\over\partial X_m}
+(\la{\bf v}_1\ra_h\cdot\nabla_{\bf x}){\bf H}$$
\begI
-2\eta(\nabla_{\bf x}\cdot\nabla_{\bf X}){\bf h}_0
-\nabla_{\bf X}\times({\bf v}_0\times{\bf H}+{\bf V}\times{\bf h}_0)
-\nabla_{\bf x}\times({\bf v}_0\times{\bf h}_0),
\endI{MorderNh}
$${\cal M}^\theta(\bom'_1,\theta_1)=-\sum_{k=1}^K\sum_{m=1}^2{\cal L}^\theta
(\nabla_{\bf x}s^v_{m,k},0)\,{\partial c_{0,k}\over\partial X_m}
+(\la{\bf v}_1\ra_h\cdot\nabla_{\bf x})\Theta$$
\begI
-2\kappa(\nabla_{\bf x}\cdot\nabla_{\bf X})\theta_0
+({\bf V}\cdot\nabla_{\bf X})\theta_0+({\bf v}_0\cdot\nabla_{\bf x})\theta_0.
\endL{MorderNtheta}

Solvability condition \rf{solveomega} is clearly verified for (\theMorderN),
and ${\bf v}'_1$ can be determined from \rf{vprime}. Condition \rf{solveh}
yields expression \rf{Malpha} for $\la{\bf h}_1\ra_h$ (see Section 6).

\underline{Step $5^\circ$ for $n=1$.} Scalar multiplying (\theMorderN) by
${\bf S}^{*\cdt}_j$, $j=1,2$, using their biorthogonality to
${\bf S}^\cdt_k$ and normalisation \rf{norm125}, and employing
\rf{eq23}, \rf{avevI} and the identity
$$-{\partial{\bf S}^v_k\over\partial t}
+\nu\nabla^2{\bf S}^v_k+{\bf V}\times{\bf S}^\omega_k+{\bf S}^v_k\times\bOm
-{\bf H}\times(\nabla\times{\bf S}^h_k)-{\bf S}^h_k\times(\nabla\times{\bf H})
+\tau{\bf S}^v_k\times{\bf e}_3+\beta S^\theta_k{\bf e}_3=\nabla S^p_k$$
obtained by ``uncurling'' the vorticity component of \rf{eq20}, we find
\begin{equation}
\lad{\bf v}_1\rad_h\!=\sum_{k=1}^K\!\left(\,\sum_{m=1}^2\!\left(
\bbeta_{1,m,k}{\partial c_{0,k}\over\partial X_m}
+\sum_{j=1}^2\sum_{i=1}^j\bbeta_{3,i,j,m,k}{\partial^3\nabla^{-2}_{\bf X}c_{0,k}\over\partial X_m\partial X_j\partial X_i}\right)
\!+\!\sum_{m=1}^k\bbeta_{2,m,k}c_{0,m}c_{0,k}\right)\!,
\label{avev1}\end{equation}
where it is denoted
$$\bbeta_{\cdtx}\equiv(\lad{\bf E}^\cdt_{\cdtx}\cdot{\bf S}^{*\cdt}_1\rad,
\lad{\bf E}^\cdt_{\cdtx}\cdot{\bf S}^{*\cdt}_2\rad,0),$$
and ${\bf E}^\cdt_\cdt$ are 7-dimensional vector fields:
$${\bf E}^\cdt_{1,m,k}\equiv{\cal L}({\bf 0},\nabla_{\bf x}s^v_{m,k},\nabla_{\bf x}s^h_{m,k},0)
-\lpar\lbdb\int_0^t\bal^\omega_{m,k}\,{\rm d}t\rbdb\cdot\nabla_{\bf x}\rpar
(\bOm,{\bf H},\Theta)$$
$$+\left(2\nu{\partial{\bf S}^\omega_k\over\partial x_m}\right.
+\nabla_{\bf x}\times\lpar{\bf V}\times({\bf e}_m\times{\bf S}_k^v)
-{\bf H}\times({\bf e}_m\times{\bf S}_k^h)\rpar
+{\bf e}_m\times\nabla_{\bf x}S_k^p+\tau{(S_k^v)}_m{\bf e}_3,$$
$$\left.2\eta{\partial{\bf S}^h_k\over\partial x_m}
+{\bf e}_m\times({\bf V}\times{\bf S}^h_k+{\bf S}^v_k\times{\bf H}),\quad
2\kappa{\partial S^\theta_k\over\partial x_m}-V_mS^\theta_k \right),$$
$${\bf E}^\cdt_{2,m,k}\equiv\rho_{m,k}\lpar\nabla_{\bf x}\times\left({\bf S}^v_k\times{\bf S}^\omega_m
-{\bf S}^h_k\times(\nabla_{\bf x}\times{\bf S}^h_m)+{\bf S}^v_m\times{\bf S}^\omega_k
-{\bf S}^h_m\times(\nabla_{\bf x}\times{\bf S}^h_k)\right)\!,$$
$$\nabla_{\bf x}\times\left({\bf S}^v_k\times{\bf S}^h_m+{\bf S}^v_m\times{\bf S}^h_k\right),
\ \ -({\bf S}^v_k\cdot\nabla_{\bf x})S^\theta_m-({\bf S}^v_m\cdot\nabla_{\bf x})S^\theta_k\rpar,$$
$${\bf E}^\cdt_{3,i,j,m,k}\equiv\rho_{i,j}\left(\lpar\lbdb\int_0^t{(\bal^\omega_{m,k})}_j\,{\rm d}t\rbdb{\bf e}_i
+\lbdb\int_0^t{(\bal^\omega_{m,k})}_i\,{\rm d}t\rbdb{\bf e}_j\rpar\cdot
\nabla_{\bf x}\right)(\bOm,{\bf H},\Theta),$$
$\rho_{m,k}\equiv1$ for $m<k$, $\rho_{m,k}\equiv1/2$ for $m=k$, and
$\rho_{m,k}\equiv0$ for $m>k$. If the perturbed CHM regime ${\bf V,H},\Theta$
possesses a symmetry considered in Zh08, vector fields ${\bf E}^\cdt_{\cdtx}$
for $k\le 4$ are symmetric, ${\bf S}^{*\cdt}_k$ for $k=1,2$ are antisymmetric,
and thereby $\lad{\bf v}_1\rad_h=\bf 0$, if ${\bf V,H},\Theta$ is a steady CHM state.

If the CHM regime ${\bf V,H},\Theta$ is unsteady, scalar multiplication
of (\theMorderN) by ${\bf S}^{*\cdt}_5$, with the use of its orthogonality to
${\bf S}^\cdt_k$, $k=1,2$ and normalisation \rf{norm125}, yields
\begin{equation}
\sum_{k=1}^K\!\left(\,\sum_{m=1}^2\!\left(
\beta'_{1,m,k}{\partial c_{0,k}\over\partial X_m}
+\sum_{j=1}^2\sum_{i=1}^j\beta'_{3,i,j,m,k}{\partial^3\nabla^{-2}_{\bf X}c_{0,k}\over\partial X_m\partial X_j\partial X_i}\right)
\!+\!\sum_{m=1}^k\beta'_{2,m,k}c_{0,m}c_{0,k}\right)=0,
\label{eqtimedep}\end{equation}
where
$$\beta'_{\cdtx}=\lad{\bf E}^\cdt_{\cdtx}\cdot{\bf S}^{*\cdt}_5\rad.$$
All the solvability conditions for (\theMorderN) have been satisfied,
and thus a solution can be constructed.

\underline{Step $6^\circ$ for $n=1$.}
Let $\widehat{\cal P}({\bf a}^\omega,{\bf a}^h,a^\theta)$ denote the projection
of a vector field\break$({\bf a}^\omega,{\bf a}^h,a^\theta)$ to the subspace,
orthogonal to $\ker{\cal M}^*$ (in the sense of the scalar product
$\lad({\bf a}_1^\omega,{\bf a}_1^h,a_1^\theta)\cdot({\bf a}_2^\omega,{\bf a}_2^h,a_2^\theta)\rad$~).
If the CHM state ${\bf V,H},\Theta$ is unsteady and symmetric, application of
$\widehat{\cal P}$ amounts to projecting out the ${\bf S}^{*\cdt}_5$ component,
since the right-hand sides in (\theMorderN) are symmetric, and
${\bf S}^{*\cdt}_5$ is the only symmetric eigenmode in $\ker{\cal M}^*$.
In view of relations \rf{eq23} and \rf{avevI},
and due to linearity of this problem, it has the following solution:
$$(\bom'_1,{\bf v}'_1,{\bf h}'_1,\theta_1)
=\bxi_1^\cdt+\left.\sum_{k=1}^K\right({\bf S}^\cdt_kc_{1,k}\phantom{\partial\over\partial}$$
\begin{equation}
\left.+\sum_{m=1}^2\left({\bf G}^\cdt_{m,k}{\partial c_{0,k}\over\partial X_m}
+\sum_{j=1}^2\sum_{i=1}^j{\bf Y}^\cdt_{i,j,m,k}{\partial^3\nabla^{-2}_{\bf X}c_{0,k}\over\partial X_m\partial X_j\partial X_i}\right)
+\sum_{m=1}^k{\bf Q}^\cdt_{m,k}c_{0,k}c_{0,m}\right)\!.
\label{eq43}\end{equation}

Here ${\bf G}^\cdt_{m,k}=({\bf G}^\omega_{m,k},{\bf G}^v_{m,k},{\bf G}^h_{m,k},G^\theta_{m,k})$
solve \underline{\it auxiliary problems of type II}:
\begB
{\cal M}({\bf G}^\cdt_{m,k})=\widehat{\cal P}{\bf E}^\cdt_{1,m,k},
\endB{eq44p1}{auxII}\begI
\nabla_{\bf x}\times{\bf G}^v_{m,k}={\bf G}^\omega_{m,k},
\endI{eq44p3}\begI
\nabla_{\bf x}\cdot{\bf G}^\omega_{m,k}=
\nabla_{\bf x}\cdot{\bf G}^v_{m,k}=\nabla_{\bf x}\cdot{\bf G}^h_{m,k}=0.
\endL{eq44sol}

Vector fields ${\bf Q}^\cdt_{m,k}=({\bf Q}^\omega_{m,k},{\bf Q}^v_{m,k},{\bf Q}^h_{m,k},Q^\theta_{m,k})$
solve \underline{\it auxiliary problems of type III}:
\begB
{\cal M}^\omega({\bf Q}^\cdt_{m,k})=\widehat{\cal P}{\bf E}^\cdt_{2,m,k},
\endB{eq46p1}{auxIII}\begI
\nabla_{\bf x}\times{\bf Q}^v_{m,k}={\bf Q}^\omega_{m,k},
\endI{eq46p2}\begI
\nabla_{\bf x}\cdot{\bf Q}^\omega_{m,k}=\nabla_{\bf x}\cdot{\bf Q}^v_{m,k}=
\nabla_{\bf x}\cdot{\bf Q}^h_{m,k}=0
\endL{eq49}
(${\bf Q}^\cdt_{m,k}=\bf 0$ for $m>k$).

Vector fields ${\bf Y}^\cdt_{i,j,m,k}=({\bf Y}^\omega_{i,j,m,k},
{\bf Y}^v_{i,j,m,k},{\bf Y}^h_{i,j,m,k},Y^\theta_{i,j,m,k})$
solve \underline{\it auxiliary problems} \underline{\it of type IV}:
\begB
{\cal M}({\bf Y}^\cdt_{i,j,m,k})=\widehat{\cal P}{\bf E}^\cdt_{3,i,j,m,k},
\endB{eq50p1}{auxIV}\begI
\nabla_{\bf x}\times{\bf Y}^v_{i,j,m,k}={\bf Y}^\omega_{i,j,m,k},
\endI{eq50p2}\begI
\nabla_{\bf x}\cdot{\bf Y}^\omega_{i,j,m,k}=
\nabla_{\bf x}\cdot{\bf Y}^v_{i,j,m,k}=\nabla_{\bf x}\cdot{\bf Y}^h_{i,j,m,k}=0
\endL{eq52}
(${\bf Y}^\cdt_{i,j,m,k}=\bf 0$ for $i>j$).

Finally, $\bxi^\cdt_1=(\bxi^\omega_1,\bxi^v_1,\bxi^h_1,\xi^\theta_1)$ solve the problem
$${\cal L}^\omega(\bxi^\omega_1,\bxi^v_1,\bxi^h_1,\xi^\theta_1)
=-2\nu(\nabla_{\bf x}\cdot\nabla_{\bf X})\bxi^\omega_0
-\nabla_{\bf X}\times\lpar{\bf V}\times\bxi^\omega_0+\bxi^v_0\times\bOm$$
$$-{\bf H}\times(\nabla_{\bf x}\times\bxi^h_0)-\bxi^h_0\times(\nabla_{\bf x}\times{\bf H})\rpar
-\nabla_{\bf x}\times\lpar{\bf v}_0\times\bxi^\omega_0+\bxi^v_0\times(\bom_0-\bxi^\omega_0)$$
\begB
-{\bf H}\times(\nabla_{\bf X}\times\bxi^h_0)
-{\bf h}_0\times(\nabla_{\bf x}\times\bxi^h_0)
-\bxi^h_0\times(\nabla_{\bf x}\times({\bf h}_0-\bxi^h_0))\rpar
-\beta\nabla_{\bf X}\xi^\theta_0\times{\bf e}_3,
\endB{eq53p1}{IIxi}
$${\cal L}^h(\bxi^v_1,\bxi^h_1)=-2\eta(\nabla_{\bf x}\cdot\nabla_{\bf X})\bxi^h_1
-\nabla_{\bf X}\times(\bxi^v_0\times{\bf H}+{\bf V}\times\bxi^h_0)$$
\begI
-\nabla_{\bf x}\times({\bf v}_0\times\bxi^h_0+\bxi^v_0\times({\bf h}_0-\bxi^h_0)),
\endI{eq53p5}\begI
{\cal L}^\theta(\bxi^v_1,\xi^\theta_1)
=-2\kappa(\nabla_{\bf x}\cdot\nabla_{\bf X})\xi^\theta_0
+({\bf V}\cdot\nabla_{\bf X})\xi^\theta_0+({\bf v}_0\cdot\nabla_{\bf x})\xi^\theta_0
+(\bxi^v_0\cdot\nabla_{\bf x})(\theta_0-\xi^\theta_0),
\endI{eq53p7}\begI
\nabla_{\bf x}\times\bxi^v_1-\bxi^\omega_1=-\nabla_{\bf X}\times\bxi^v_0,
\endI{eq53p3}\begI
\nabla_{\bf x}\cdot\bxi_1^\omega=-\nabla_{\bf X}\cdot\bxi_0^\omega,\quad
\nabla_{\bf x}\cdot\bxi_1^v=-\nabla_{\bf X}\cdot\bxi_0^v,\quad
\nabla_{\bf x}\cdot\bxi_1^h=-\nabla_{\bf X}\cdot\bxi_0^h.
\endL{eq53sol}

${\bf G}^\cdt_{m,k},{\bf Q}^\cdt_{m,k},{\bf Y}^\cdt_{i,j,m,k}$ and
$\bxi^\cdt_1$ must satisfy the boundary conditions \xrf{eq2SF}{eq2EC} and
\rf{eq4T}. Spatial means
of their horizontal flow components and of vertical vorticity components must
vanish, as well as spatio-temporal means of horizontal magnetic components.

Together, \rf{eq44sol}, \rf{eq49}, \rf{eq52} and \rf{eq53sol} are equivalent
to \xrf{eq17p2}{eq17p5} for $n=1$. It suffices to impose
these conditions for the vorticity and magnetic parts of a solution at $t=0$
(this can be deduced taking the divergence of the vorticity and magnetic
equations of the problems (\theauxII)--(\theIIxi)~). Equations \rf{eq44p3},
\rf{eq46p2}, \rf{eq50p2} and \rf{eq53p3} are equivalent to \rf{eq30r} for $n=1$.

The spatial means of the initial conditions for these problems
can be determined integrating the equations for vorticity and magnetic
perturbations and considering the spatio-temporal means of the results:
\begin{equation}
\phan\la{\bf G}^h_{m,k}\ra_h\right|_{t=0}
=-\ladb{\bf e}_m\times\int_0^t\bal^h_k\,{\rm d}t\radb_{\!h}\!,
\label{EQbegin0}\end{equation}
$$\phan\la\bxi^h_1\ra_h\right|_{t=0}=-\ladb\int_0^t\nabla_{\bf X}\times\txi{h}\,{\rm d}t\radb_{\!h}$$
(the means exist by the assumption that the magnetic $\alpha-$effect is
insignificant); at any $t\ge0$
\begin{equation}
\la{\bf G}^\omega_{m,k}\ra_v=\la{\bf Q}^\omega_{m,k}\ra_v
=\la{\bf Y}^\omega_{i,j,m,k}\ra_v=\la\bxi^\omega_1\ra_v
=\la{\bf Q}^h_{m,k}\ra_h=\la{\bf Y}^h_{i,j,m,k}\ra_h={\bf 0}.
\label{EQbegin2}\end{equation}

Averaging the magnetic part of \rf{eq43}, we obtain
$$\phan\lad{\bf h}_1\rad_h\right|_{T=0}
\phan=\la{\bf h}_1\ra_h\right|_{t=0}\phan-\la\bxi^h_1\ra_h\right|_{t=0}
-\sum_{k=1}^K\sum_{m=1}^2
\phan\la{\bf G}^h_{m,k}\ra_h\right|_{t=0}{\partial\lad c_{0,k}\rad\over\partial X_m}\phan\!\right|_{T=0}.$$
Now initial conditions for the problem (\theIIxi) can be determined
from \rf{eq43} and (\thevhomone) at $t=0$.

The choice of initial data must ensure that all solutions to the auxiliary
problems are globally bounded with their derivatives. Any solution is globally
bounded for any smooth initial conditions, if the perturbed CHM state
${\bf V,H},\Theta$ is space-periodic and linearly stable to
small-scale perturbations (see Appendix B in Zh08). Modification of initial
conditions for ${\bf G}^\cdt_{m,k}$, ${\bf Q}^\cdt_{m,k}$ and
${\bf Y}^\cdt_{i,j,m,k}$ into other permissible ones implies the respective
changes in initial conditions for $\bxi^\cdt_1$. These changes must belong
to the stable manifold of the perturbed
CHM state ${\bf V,H},\Theta$, so that the resultant changes in solutions to the
auxiliary problems and in $\bxi^\cdt_1$ decay exponentially in fast time.

Since $\bxi^\cdt_0$ decay exponentially in fast time together with derivatives
(see Section 4), the right-hand sides of equations (\theIIxi) also
do. If the CHM state ${\bf V,H},\Theta$ is linearly stable to small-scale
perturbations, this implies that $\bxi^\cdt_1$ decay exponentially,
and any changes in ${\bf G}^\cdt_{m,k}$, ${\bf Q}^\cdt_{m,k}$ and
${\bf Y}^\cdt_{i,j,m,k}$ due to a permissible variation of the initial data for
${\bf S}^\cdt_k$ also exponentially decay (see Zh08, Appendix B). Otherwise,
exponential decay
of $\bxi^\cdt_1$ must be ensured by an appropriate choice of the initial data.

If the CHM regime ${\bf V,H},\Theta$ is steady or periodic in time and periodic
in spatial variables, generically one can find ${\bf S}^\cdt_k$,
${\bf G}^\cdt_{m,k}$, ${\bf Q}^\cdt_{m,k}$ and ${\bf Y}^\cdt_{i,j,m,k}$,
which are steady or have the same periodicity, respectively (see Section 4
in Zh08). If the perturbed CHM regime has a symmetry of
the kind considered in Zh08, ${\bf S}^\cdt_k$ are antisymmetric sets, except
for ${\bf S}^\cdt_5$, which is a symmetric set of vector fields for non-steady
regimes. Consequently,
${\bf G}^\cdt_{m,k}$, ${\bf Q}^\cdt_{m,k}$ and ${\bf Y}^\cdt_{i,j,m,k}$
are then symmetric sets for $k<5$ (by construction the antisymmetric part
of any permissible solution to the problems (\theauxII)--(\theauxIV) for $k<5$
exponentially decays and hence is irrelevant), as well as ${\bf Q}^\cdt_{m,5}$;
${\bf G}^\cdt_{m,5}$ and ${\bf Y}^\cdt_{i,j,m,5}$ are antisymmetric sets.
For symmetric sets conditions \rf{EQbegin0} and \rf{EQbegin2} are automatically
satisfied (except for the mean of the vertical component of vorticity,
vanishing of which is not implied by parity invariance of the perturbed state).
If the perturbed CHM state is steady or its symmetry is without a time shift,
the right-hand sides of auxiliary problems (\theauxIV) are zero and hence
${\bf Y}^\cdt_{i,j,m,k}=\bf 0$.

\mi{\bf 8. Order $\varepsilon^2$ and $\varepsilon^3$ equations: amplitude equations}

\mi
\underline{Step $1^\circ$ for $n=2$.} Combining the flow velocity component
of \rf{eq43}, \rf{avev1}, \rf{vone} and \rf{avevI}, we find
$${\bf v}_1=\bxi_1^v+\sum_{k=1}^K\left({\bf S}^v_kc_{1,k}
+\sum_{m=1}^2\left(\widetilde{\bf G}^v_{m,k}{\partial c_{0,k}\over\partial X_m}
+\sum_{j=1}^2\sum_{i=1}^j\widetilde{\bf Y}^v_{i,j,m,k}{\partial^3\nabla^{-2}_{\bf X}c_{0,k}\over\partial X_m\partial X_j\partial X_i}\right)\right.$$
\begB
+\left.\sum_{m=1}^k\widetilde{\bf Q}^v_{m,k}c_{0,k}c_{0,m}\right),
\endB{v1full}{vhfull}

\noindent
where
$$\widetilde{\bf G}^v_{m,k}={\bf G}^v_{m,k}+\bbeta_{1,m,k}+\lbdb\int_0^t\bal^\omega_{m,k}\,{\rm d}t\rbdb
+\nabla_{\bf x}s^v_{m,k},$$
$$\widetilde{\bf Q}^v_{m,k}={\bf Q}^v_{m,k}+\bbeta_{2,m,k},$$
$$\widetilde{\bf Y}^v_{i,j,m,k}={\bf Y}^v_{i,j,m,k}
+\rho_{i,j}\lpar\bbeta_{3,i,j,m,k}-\lbdb\int_0^t\!(\bal^\omega_{m,k})_j\,{\rm d}t\!\rbdb{\bf e}_i
+\bbeta_{3,j,i,m,k}-\lbdb\int_0^t\!(\bal^\omega_{m,k})_i\,{\rm d}t\!\rbdb{\bf e}_j\rpar.$$
Similarly,
$${\bf h}_1=\bxi_1^h+\sum_{k=1}^K\left({\bf S}^h_kc_{1,k}
+\sum_{m=1}^2\left(\widetilde{\bf G}^h_{m,k}{\partial c_{0,k}\over\partial X_m}
+\sum_{j=1}^2\sum_{i=1}^j{\bf Y}^h_{i,j,m,k}{\partial^3\nabla^{-2}_{\bf X}c_{0,k}\over\partial X_m\partial X_j\partial X_i}\right)\right.$$
\begI
+\left.\sum_{m=1}^k{\bf Q}^h_{m,k}c_{0,k}c_{0,m}\right),
\endL{h1full}
where
$$\widetilde{\bf G}^h_{m,k}={\bf G}^h_{m,k}+\nabla_{\bf x}s^h_{m,k}.$$

Averaging (A.4) for $n=3$ in fast time, substituting \rf{avom1}, \rf{eq34},
(\thevhfull) and the flow and magnetic components of \rf{eq23}, and recalling
that $\bxi^\cdt_0$ and $\bxi^\cdt_1$ decay exponentially, we obtain
$$\nabla_{\bf X}\times\sum_{k=1}^K\sum_{n=1}^2{\partial\over\partial X_n}\left(
\sum_{m=1}^2{\partial\over\partial X_m}
\left({\bf D}^v_{n,m,k}c_{0,k}+\sum_{j=1}^2\sum_{i=1}^j{\bf d}^v_{n,i,j,m,k}
{\partial^2\over\partial X_i\partial X_j}\nabla^{-2}_{\bf X}c_{0,k}\right)\right.$$
\begin{equation}
\left.+\sum_{m=1}^k{\bf A}^v_{n,m,k}c_{0,m}c_{0,k}\right)={\bf 0},
\label{eq59}\end{equation}
where
$${\bf D}^v_{n,m,k}=\lad-V_n\widetilde{\bf G}^v_{m,k}-{\bf V}{(\widetilde G^v_{m,k})}_n
+H_n\widetilde{\bf G}^h_{m,k}+{\bf H}{(\widetilde G^h_{m,k})}_n\rad_h,$$
$${\bf d}^v_{n,i,j,m,k}=\lad-V_n\widetilde{\bf Y}^v_{i,j,m,k}-{\bf V}{(\widetilde Y^v_{i,j,m,k})}_n
+H_n{\bf Y}^h_{i,j,m,k}+{\bf H}{(Y^h_{i,j,m,k})}_n\rad_h,$$
$${\bf A}^v_{n,m,k,j}=\lad-V_n\widetilde{\bf Q}^v_{m,k}-{\bf V}{(\widetilde{Q}^v_{m,k})}_n
+H_n{\bf Q}^h_{m,k}+{\bf H}{(Q^h_{m,k})}_n-{(S^v_k)}_n{\bf S}^v_m+{(S^h_k)}_n{\bf S}^h_m\rad_h.$$

\underline{Step $2^\circ$ for $n=2$.}
Averaging the horizontal component of (A.2) for $n=2$ in fast variables,
substituting (\thevhfull) and flow and magnetic components of \rf{eq23},
recalling that $\bxi^\cdt_0$ and $\bxi^\cdt_1$ decay exponentially,
and taking into account relations \rf{eq11p2}, \rf{eq37} and the boundary
conditions for $\bf V,\ H$, ${\bf v}_n$ and ${\bf h}_n$, we find
$$-{\partial\over\partial T}\lad{\bf h}_0\rad_h+\eta\nabla^2_{\bf X}\lad{\bf h}_0\rad_h
+\nabla_{\bf X}\times\sum_{k=1}^K\left(\,\sum_{m=1}^2{\partial\over\partial X_m}\right({\bf D}^h_{m,k}c_{0,k}$$
\begin{equation}
+\sum_{j=1}^2\sum_{i=1}^j{\bf d}^h_{i,j,m,k}{\partial^2\over\partial X_i\partial X_j}\nabla^{-2}_{\bf X}c_{0,k}\left)
+\sum_{m=1}^k{\bf A}^h_{m,k}c_{0,k}c_{0,m}\right)={\bf 0},
\label{eq62}\end{equation}
where
$${\bf D}^h_{m,k}=\lad{\bf V}\times\widetilde{\bf G}^h_{m,k}-{\bf H}\times\widetilde{\bf G}^v_{m,k}\rad_v,$$
$${\bf d}^h_{i,j,m,k}=\lad{\bf V}\times{\bf Y}^h_{i,j,m,k}-{\bf H}\times\widetilde{\bf Y}^v_{i,j,m,k}\rad_v,$$
$${\bf A}^h_{m,k}=\lad{\bf V}\times{\bf Q}^h_{m,k}-{\bf H}\times\widetilde{\bf Q}^v_{m,k}
+{\bf S}^v_k\times{\bf S}^h_m\rad_v.$$

${\bf D}^\cdt$ are coefficients of the second-order operators representing
the so-called {\it aniso\-tropic combined eddy diffusivity correction}.
${\bf d}^\cdt$ are coefficients of pseudodifferential operators,
formally also of the second order, which can be regarded as representing an
unconventional {\it non-local anisotropic combined eddy diffusivity
correction}. All ${\bf d}^\cdt=\bf 0$, if the perturbed CHM state is steady or
possesses a symmetry considered in Zh08 without a time shift.
${\bf A}^\cdt$ are coefficients of quadratic terms
representing the so-called {\it combined eddy advection correction}.

Equations \rf{eq59} and \rf{eq62} are solvability conditions for the systems
(A.1)--(A.3) for $n=2$ and 3. Together with \rf{eqtimedep}, if the CHM
regime ${\bf V,H},\Theta$ is unsteady, they constitute a closed system of
equations for the leading terms of mean perturbations (note that
$c_{0,k+2}=\lad{\bf h}_0\rad_k$, $k=1,2$, since
$\la{\bf S}^h_{k+2}\ra_h={\bf e}_k$). The vertical component of \rf{eq62}
and the horizontal component of \rf{eq59} vanish identically (i.e.,
\rf{eq62} represents two scalar equations and \rf{eq59} just one).

Unlike for forced CHM systems, the system of amplitude equations
is mixed, involving both evolutionary and non-evolutionary equations
in slow variables. Equation \rf{eq62} for the mean magnetic field
perturbation is evolutionary. It preserves solenoidality of the mean magnetic
perturbation (condition \rf{eq17p1} for $n=0$) in slow spatial variables,
which therefore becomes just a constraint for the initial condition. Equation
\rf{eq59} resulting from the vorticity equation does not bear any similarity
with the original equation -- both the derivative in slow time and molecular
diffusivity operator are absent. Equation \rf{eqtimedep}, emerging if
the CHM regime ${\bf V,H},\Theta$ is unsteady, is also non-evolutionary,
as well as the condition of solenoidality in slow variables
\rf{eq17p1} for the mean flow $\lad{\bf v}_1\rad_h$ \rf{avev1}.
The system of amplitude equations is still underdetermined, if the CHM
state is steady and possesses the symmetries considered in Zh08, resulting in
$\lad{\bf v}_1\rad_h=\bf 0$. The missing equation is then the condition
of solenoidality in slow variables of the mean flow $\la{\bf v}_2\ra_h$.

\mi{\bf 9. Mean flow perturbation $\la{\bf v}_2\ra_h$
for a symmetric steady CHM state}

\mi
In this section we calculate $\la{\bf v}_2\ra_h$ for a steady CHM state
${\bf V,H},\Theta$, which is parity-invariant or symmetric about a vertical
axis. In this case, in view of
\rf{avom1}, \rf{eq39}, \rf{avevI}, \rf{avev1} and \rf{eq19} for $n=0$ and 1,
$$\la{\bf v}_0\ra_h=\la{\bf v}_1\ra_h=\la\bom_0\ra_v=\la\bom_1\ra_v=\la\bom_2\ra_v=\bf 0.$$
This is not an independent calculation -- we advance by two steps in solution
of the system (A.1)--(A.3) for $n=2$, following the general procedure
for treatment of the systems in the hierarchy that is outlined in the end of
Section 3 (although the solution itself is not required for our purposes).

\underline{Step $3^\circ$ for $n=2$.}
We need to apply solvability conditions \rf{solves} to
equations (A.1)--(A.3) for $n=2$. To do this, we transform the system to
fit the form of the problem (\theMsys) by isolating the gradient
parts of $\bom_2,{\bf v}_2$ and ${\bf h}_2$, and the mean part of ${\bf v}_2$
using \rf{eq30r}, \xrf{eq17p2}{eq17p5} for $n=2$ and \rf{eq43}:
\begB
\bom_2\!=\sum_{k=1}^K\sum_{n=1}^2{\partial\over\partial X_n}\left(\,\sum_{m=1}^2
{\partial c_{0,k}\over\partial X_m}\,{\bf e}_n\times\widetilde{\bf G}^v_{m,k}
\!+\!\sum_{m=1}^Kc_{0,m}c_{0,k}\,{\bf e}_n\times{\bf Q}^v_{m,k}\right)\!+\bom'_2,
\endB{omtwo}{vhomtwo}\begI
{\bf v}_2\!=\la{\bf v}_2\ra_h\!+\!\sum_{k=1}^K\sum_{n=1}^2{\partial\over\partial X_n}
\left(\,\sum_{m=1}^2{\partial c_{0,k}\over\partial X_m}\,\nabla_{\bf x}g^v_{n,m,k}
\!+\!\sum_{m=1}^Kc_{0,m}c_{0,k}\nabla_{\bf x}q^v_{n,m,k}\right)\!+\!{\bf v}'_2,
\endI{vtwo}\begI
{\bf h}_2=\sum_{k=1}^K\sum_{n=1}^2{\partial\over\partial X_n}\left(\,\sum_{m=1}^2
{\partial c_{0,k}\over\partial X_m}\,\nabla_{\bf x}g^h_{n,m,k}
\!+\!\sum_{m=1}^Kc_{0,m}c_{0,k}\nabla_{\bf x}q^h_{n,m,k}\right)\!+\!{\bf h}'_2,
\endL{htwo}
where
$$\nabla_{\bf x}\cdot\bom'_2=\nabla_{\bf x}\cdot{\bf v}'_2
=\nabla_{\bf x}\cdot{\bf h}'_2=0,\qquad\la\bom'_2\ra_v=\la{\bf v}'_2\ra_h={\bf 0},\qquad
\nabla_{\bf x}\times{\bf v}'_2=\bom'_2,$$
$g^\cdt_{n,m,k}({\bf x},t)$ and $q^\cdt_{n,m,k}({\bf x},t)$
are globally bounded solutions to Neumann problems
$$\nabla^2_{\bf x}\,g^v_{n,m,k}=-(\widetilde{G}^v_{m,k})_n,\quad
\left.{\partial g^v_{n,m,k}\over\partial x_3}\right|_{x_3=\pm L/2}=0;$$
$$\nabla^2_{\bf x}\,g^h_{n,m,k}=-(\widetilde{G}^h_{m,k})_n,\quad
\left.{\partial g^h_{n,m,k}\over\partial x_3}\right|_{x_3=\pm L/2}=0;$$
$$\nabla^2_{\bf x}\,q^v_{n,m,k}=-(Q^v_{m,k})_n,\quad
\left.{\partial q^v_{n,m,k}\over\partial x_3}\right|_{x_3=\pm L/2}=0;$$
$$\nabla^2_{\bf x}\,q^h_{n,m,k}=-(Q^h_{m,k})_n,\quad
\left.{\partial q^h_{n,m,k}\over\partial x_3}\right|_{x_3=\pm L/2}=0,$$
and in agreement with the previously used notation (recall that $\bbeta_\cdt=\bf0$
in \rf{avev1} for a perturbation of a symmetric CHM steady state)
$$\widetilde{\bf G}^\cdt_{m,k}=({\bf G}^\omega_{m,k}+{\bf e}_m\times{\bf S}^v_k,
{\bf G}^v_{m,k}+\nabla_{\bf x}s^v_{m,k},
{\bf G}^h_{m,k}+\nabla_{\bf x}s^h_{m,k},G^\theta_{m,k}).$$
The fields $\bom'_2,{\bf v}'_2$ and ${\bf h}'_2$ satisfy the boundary
conditions for vorticity, flow and magnetic field, respectively.

\underline{Step $4^\circ$ for $n=2$.}
Substituting (\thevhomtwo), \rf{eq23} where $\bxi_0^\cdt=\bf 0$, and
$$(\bom_1,{\bf v}_1,{\bf h}_1,\theta_1)=\sum_{k=1}^K\left({\bf S}^\cdt_kc_{1,k}
+\sum_{m=1}^2\widetilde{\bf G}^\cdt_{m,k}{\partial c_{0,k}\over\partial X_m}
+\sum_{m=1}^K{\bf Q}^\cdt_{m,k}c_{0,k}c_{0,m}\right)$$
(a consequence of \rf{eq43} and (\thevhomone), where $\la{\bf v}_1\ra_h=\bf 0$)
into equations (A.1)--(A.3) for $n=2$, noting that $c_{1,k}$
enter these equations only via terms which are symmetric sets, scalar
multiplying the resultant equations by ${\bf S}^{*\cdt}_j$, $j=1,2$ and using
their biorthogonality to ${\bf S}^\cdt_k$ and normalisation \rf{norm125}, we find
$$\la{\bf v}_2\ra_h=\sum_{k=1}^K\left(\,\sum_{m=1}^2\sum_{n=1}^2
\bbeta''_{1,n,m,k}{\partial^2 c_{0,k}\over\partial X_m\partial X_n}
+\sum_{m=1}^K\sum_{n=1}^2\bbeta''_{2,n,m,k}c_{0,m}{\partial c_{0,k}\over\partial X_n}\right.$$
\begin{equation}
+\left.\sum_{m=1}^K\sum_{n=1}^K\bbeta''_{3,n,m,k}c_{0,k}c_{0,m}c_{0,n}\right).
\label{avev2}\end{equation}
Here it is denoted
$$\bbeta''_{i,n,m,k}\equiv(\la\widetilde{\bf E}^\cdt_{i,n,m,k}\cdot{\bf S}^{*\cdt}_1\ra,
\la\widetilde{\bf E}^\cdt_{i,n,m,k}\cdot{\bf S}^{*\cdt}_2\ra,0),$$
where $\widetilde{\bf E}^\cdt_\cdt$ are 7-dimensional vector fields:
$$\widetilde{\bf E}^\cdt_{1,n,m,k}\equiv{\cal L}({\bf e}_n\times\widetilde{\bf G}^v_{m,k},
\nabla_{\bf x}\,g^v_{n,m,k},\nabla_{\bf x}\,g^h_{n,m,k},0)
+\left(\,2\nu{\partial\over\partial x_n}\widetilde{\bf G}^\omega_{m,k}\right.$$
$$+\,{\bf e}_n\times\lpar{\bf V}\times\widetilde{\bf G}^\omega_{m,k}
+\widetilde{\bf G}^v_{m,k}\times\bOm
-{\bf H\times(\nabla_x}\times\widetilde{\bf G}^h_{m,k})
-\widetilde{\bf G}^h_{m,k}\times(\nabla_{\bf x}\times{\bf H})$$
$$-{\bf H}\times({\bf e}_m\times{\bf S}^h_k)-\beta G^\theta_{m,k}{\bf e}_3\rpar
-\nabla_{\bf x}\times({\bf H}\times({\bf e}_n\times\widetilde{\bf G}^h_{m,k})),$$
$$2\eta{\partial\over\partial x_n}\widetilde{\bf G}^h_{m,k}
+{\bf e}_n\times(\widetilde{\bf G}^v_{m,k}\times{\bf H}
+{\bf V}\times\widetilde{\bf G}^h_{m,k}),\quad
\left.2\kappa{\partial\over\partial x_n}G^\theta_{m,k}-V_nG^\theta_{m,k}\right);$$

$$\widetilde{\bf E}^\cdt_{2,n,m,k}\equiv
\widetilde{\bf e}^\cdt_{n,m,k}+\widetilde{\bf e}^\cdt_{n,k,m}$$
$$+\,\lpar\nabla_{\bf x}\times(\widetilde{\bf G}^v_{n,k}\times{\bf S}^\omega_m
+{\bf S}^v_m\times\widetilde{\bf G}^\omega_{n,k}
-\widetilde{\bf G}^h_{n,k}\times(\nabla_{\bf x}\times{\bf S}^h_m)
-{\bf S}^h_m\times(\nabla_{\bf x}\times\widetilde{\bf G}^h_{n,k})),$$
$$\nabla_{\bf x}\times(\widetilde{\bf G}^v_{n,k}\times{\bf S}^h_m
+{\bf S}^v_m\times\widetilde{\bf G}^h_{n,k}),\quad
-(\widetilde{\bf G}^v_{n,k}\cdot\nabla_{\bf x})S^\theta_m
-({\bf S}^v_m\cdot\nabla_{\bf x})G^\theta_{n,k}\rpar;$$

$$\widetilde{\bf e}^\cdt_{n,m,k}\equiv{\cal L}({\bf e}_n\times{\bf Q}^v_{m,k},
\nabla_{\bf x}\,q^v_{n,m,k},\nabla_{\bf x}\,q^h_{n,m,k},0)
+\left(\,2\nu{\partial\over\partial x_n}{\bf Q}^\omega_{m,k}\right.$$
$$+\,{\bf e}_n\times\lpar{\bf V}\times{\bf Q}^\omega_{m,k}+{\bf Q}^v_{m,k}\times\bOm
-{\bf H\times(\nabla_x\times Q}^h_{m,k})-{\bf Q}^h_{m,k}\times(\nabla_{\bf x}\times{\bf H})$$
$$+\,{\bf S}^v_m\times{\bf S}^\omega_k-{\bf S}^h_m\times(\nabla_{\bf x}\times{\bf S}^h_k)
-\beta Q^\theta_{m,k}{\bf e}_3\rpar
-\nabla_{\bf x}\times({\bf H}\times({\bf e}_n\times{\bf Q}^h_{m,k})),$$
$$2\eta{\partial\over\partial x_n}{\bf Q}^h_{m,k}
+{\bf e}_n\times({\bf Q}^v_{m,k}\times{\bf H}+{\bf V}\times{\bf Q}^h_{m,k}
+{\bf S}^v_m\times{\bf S}^h_k),\quad
\left.2\kappa{\partial\over\partial x_n}Q^\theta_{m,k}-V_nQ^\theta_{m,k}\right);$$

$$\widetilde{\bf E}^\cdt_{3,n,m,k}\equiv
\lpar\nabla_{\bf x}\times({\bf Q}^v_{m,k}\times{\bf S}^\omega_n
+{\bf S}^v_n\times{\bf Q}^\omega_{m,k}
-{\bf Q}^h_{m,k}\times(\nabla_{\bf x}\times{\bf S}^h_n)
-{\bf S}^h_n\times(\nabla_{\bf x}\times{\bf Q}^h_{m,k})),$$
$$\nabla_{\bf x}\times({\bf Q}^v_{m,k}\times{\bf S}^h_n
+{\bf S}^v_n\times{\bf Q}^h_{m,k}),\quad
-\,({\bf Q}^v_{m,k}\cdot\nabla_{\bf x})S^\theta_n
-({\bf S}^v_n\cdot\nabla_{\bf x})Q^\theta_{m,k}\rpar.$$
Thus, for a steady CHM state possessing a symmetry which guarantees
insignificance of the $\alpha-$effect, the missing amplitude equation
-- the solenoidality condition for the flow \rf{avev2} -- has been expressed
in the terms of the amplitudes $c_{0,k}$.

\mi{\bf 10. Non-zero initial mean flow perturbation $\lad{\bf v}_0\rad_h$}

\mi
It has been found in Section 5 that unless the horizontal part of the mean flow
perturbation vanishes, no solution of order $\varepsilon^0$ equations
\rf{Morder1} exists in the functional space $\cal D$ (introduced in Section 4).
We discuss here the possibility of relaxing the condition that the solution
belongs to $\cal D$.

Direct substitution demonstrates that \rf{Morder1} has a solution
\begin{equation}
(\bom_0,{\bf h}_0,\theta_0)=\sum_{k=1,2}\lad{\bf v}_0\rad_k
\,t{\partial\over\partial x_k}(\bOm,{\bf H},\Theta).
\label{tgrow}\end{equation}

\pagebreak\noindent
It satisfies all the conditions for vector fields in $\cal D$, except for it
is not uniformly bounded in the fast time $t$. (Other solutions to \rf{Morder1}
can be constructed, allowing their linear growth in horizontal directions,
but they seem too unphysical.)

If growing in time solutions are regarded as acceptable, further application
of the asymptotic techniques does not require considering dependence of the
perturbation \rf{pertser} on the slow time $T$. Construction of an expansion
of large-scale solutions to the problem (\theCHMB) in power series becomes
more straightforward. At each order $\varepsilon^n$ one proceeds in two steps:\\
$1.$ Calculate $\la\bom_{n+1}\ra_v$ and $\la{\bf v}_n\ra_h$
using (A.4) at order $\varepsilon^{n+1}$.\\
$2.$ Find $\lb\bom_n\rb_v,\lb{\bf v}_n\rb_h,{\bf h}_n$ and $\theta_n$
integrating in time $t$ the system (A.1)--(A.3) at order $\varepsilon^n$.

For the sake of simplicity, suppose that the CHM state $\bOm,{\bf H},\cal T$
is steady and periodic in horizontal directions. Then the system (A.1)--(A.3)
can be solved expanding the unknown fields $\bom_n,{\bf h}_n,\theta_n$
in a series in eigenfunctions of the operator of linearisation of the original
equations, $\cal L$, which is an elliptic operator (not involving the derivative
in time). If the steady CHM state is stable to short-scale perturbations, all
eigenvalues of this operator have non-positive real parts. The coefficients
of the expansion are accessed by scalar multiplication of the system by the
biorthogonal eigenfunctions of the adjoint operator. By virtue of \rf{tgrow},
the zero-order terms $\bom_0,{\bf v}_0,{\bf h}_0,\theta_0$ grow linearly
in time. As a result, $\bom_n,{\bf v}_n,{\bf h}_n,\theta_n$ experience
a polynomial growth in time, the degree of the polynomial being $2n+1$.

Thus, free convection is apparently unstable to large-scale perturbations with
a non-zero mean flow. However, this is not a genuine instability, but rather
variation (at a given point in space) of the profile of the perturbed CHM
fields $\bOm,{\bf H},\Theta$ due to advection by the mean flow
$\lad{\bf v}_0\rad_h$. Were this mean flow constant, one could rely on the
Galilean invariance to achieve $\lad{\bf v}_0\rad_h=\bf 0$, i.e., to overcome
the difficulty by considering the problem in the comoving coordinate system.
(Note that the solution \rf{tgrow} to the order $\varepsilon^0$ problem
\rf{Morder1} represents the second term in the Taylor expansion in time of the
fields $\bOm,{\bf H},\Theta$ transported by the flow $\lad{\bf v}_0\rad_h$.)
It is conceivable that a similar transformation modifies the present weakly
stability problem into one solvable in $\cal D$, but dependence of
$\lad{\bf v}_0\rad_h$ on slow variables makes the new problem significantly
harder, than the problem considered here.

\mi{\bf 11. Conclusion}

\mi
We have investigated, in the Boussinesq approximation, weakly nonlinear
stability to large-scale perturbations of a convective hydromagnetic regime
${\bf V,H},\cal T$ in a horizontal layer rotating about a vertical axis.
We have assumed that the CHM system is free, i.e., there are no source
terms in the governing equations (\theCHMB) and boundary conditions (\theBCs).
Such system is translation invariant in space and time. We have derived
expressions for the $\alpha-$effect tensors \rf{alome} and \rf{alh}. If the
CHM regime is symmetric about a vertical axis with a time shift, or parity
invariant with a time shift, the $\alpha-$effect tensors do not contribute
to the leading order amplitude equations, i.e. the $\alpha-$effect is
insignificant in the leading order. Applying the homogenisation techniques
within the approach developed in Zh08, under the assumption that the
$\alpha-$effect is insignificant in the leading order (because of the
symmetries or otherwise) we have derived a system of amplitude equations,
\rf{eq59} and \rf{eq62}, augmented by \rf{eqtimedep} if the perturbed CHM
regime is non-steady. This system, governing the evolution of large-scale
perturbations in slow time, is closed, when considered together with
the condition of solenoidality in slow spatial variables of the leading term
of expansion of the mean flow perturbation.
In a generic $\alpha-$effect-free setup, this is the condition of solenoidality
of the flow \rf{avev1}. If the CHM state is steady and possesses a symmetry
guaranteeing insignificance of the $\alpha-$effect, $\lad{\bf v}_1\rad_h=\bf 0$
and this condition becomes trivially satisfied; then the remaining amplitude
equation is the solenoidality condition \rf{eq17p1} for the flow \rf{avev2}.

Similarly to the case of forced thermal hydromagnetic convection, which was
studied in Zh08, the system of amplitude equations involves a linear operator
of combined eddy diffusivity correction and a quadratic operator of anisotropic
combined eddy advection. Generalised diffusivity correction is described by
non-local pseudodifferential operators, formally of the second order, like the
standard diffusivity. All the operators are anisotropic. No similarity of amplitude
equations for free CHM regimes and mean-field equations for forced CHM regimes
is observed beyond this point: Because of the differences in the structure of
the kernels of the operators of linearisation in the vicinity of the perturbed
regime in the two cases, the system of amplitude equations that we have
obtained here does not involve mean flow perturbation, and it is mixed. Whereas
equation \rf{eq62} for the mean magnetic perturbation is evolutionary, like
it is for forced convection, the remaining ones, \rf{eq59} and \rf{eqtimedep},
involve neither time derivatives, nor molecular diffusivity operators.
If the CHM state is steady and possesses a symmetry guaranteeing insignificance
of the $\alpha-$effect, the solenoidality condition for the flow \rf{avev2}
reduces to a non-evolutionary third-order partial differential equation
with a cubic nonlinearity.

As already mentioned above, the neutral linear stability modes ${\bf S}_k^{\cdt,x}$
and ${\bf S}^{\cdt,t}$ exist for free CHM regimes for any combination of
physically sound boundary conditions because of the spatial and temporal
invariance of the CHM regime ${\bf V,H},\Theta$. This opens a possibility of
a similar analysis of linear or weakly nonlinear stability of CHM regimes for boundary
conditions different from the ones considered here. For instance, it is
feasible for electrically conducting fluid contained in a horizontal rotating
layer confined between half-spaces of dielectric material with isothermal no
slip boundaries, when the collection of neutral modes is the smallest one --
for generic sets of parameter values it is comprised of the three
above-mentioned modes. Note that a direct application of this approach to the
study of stability of forced convection fails for these boundary conditions.

It is possible to perform the same analysis for branches of CHM regimes
emerging from symmetric ones near Hopf and pitchfork bifurcations,
like it was done in Zh08 for forced CHM regimes. New $\alpha-$effect terms
emerge in the amplitude equations for the evolution of large-scale
perturbations in stability analysis of regimes near the point of
a symmetry-breaking Hopf bifurcation, and an additional amplitude equation
involving cubic nonlinearity emerges in stability analysis for regimes
emerging in a symmetry-breaking pitchfork bifurcation. Since the derivations,
although quite straightforward, are very technical, we have not carried out
them here.

We have not prescribed any boundary conditions for the amplitude equations,
since this would not affect our derivations. The basic requirement is that
the perturbation is globally bounded -- this is the logical foundation for
application of asymptotic methods employed here. In computations it is natural
to assume periodicity of the amplitudes in slow variables. Results of
a numerical study of large-scale perturbations under this condition, for a class
of CHM regimes symmetric about the vertical axis and stable to short-scale
perturbations, will be reported in a sequel to the present paper.

\mi{\bf Acknowledgments}

\mi
Part of this research has been carried out during my visit to the School of
Engineering, Computer Science and Mathematics, University of Exeter, UK, in
January -- April 2008. I~am grateful to the Royal Society and the University of
Exeter for their financial support. My research visits to Observatoire de la
C\^ote d'Azur were financed by the Ministry of Education of France. I was
partially supported by the grants BLAN07-2 183172 (research project OTARIE)
from Agence nationale de la recherche (France) and 07-01-92217-CNRSL\_a
from the Russian foundation for basic research.

\mi{\bf References}

\mi Dubrulle, B. and Frisch, U. Eddy viscosity of parity-invariant flow,
{\it Phys. Rev. A}, 1991, {\bf 43}, 5355--5364.

\mi Frisch, U., She, Zh.S. and Sulem, P.L. Large-scale flow driven by the
anisotropic kinetic alpha effect, {\it Physica D}, {\bf 28}, 1987, 382--392.

\mi Podvigina, O.M. Magnetic field generation by convective flows in a plane
layer, {\it Eur. Phys. J. B}, {\bf 50}, 2006, 639--652.

\mi Podvigina, O.M. Instability of flows near the onset of convection in a
rotating layer with stress-free horizontal boundaries,
{\it Geophys.~Astrophys.~Fluid Dyn.}, {\bf 102}, 2008a, 299-326.

\mi Podvigina, O.M. Magnetic field generation by convective flows
in a plane layer: the dependence on the Prandtl number,
{\it Geophys.~Astrophys.~Fluid Dyn.}, {\bf 102}, 2008b, 409--433.

\mi Podvigina, O.M. On stability of flows near the onset of convection
in a layer with stress-free boundaries,
{\it Geophys.~Astrophys.~Fluid Dyn.}, 2009, submitted.

\mi Zheligovsky, V.A. On the linear stability of spatially periodic steady
magnetohydrodynamic systems with respect to long-period perturbations,
{\it Izvestiya, Physics of the Solid Earth}, {\bf 39} (5), 2003, 409--418.

\mi Zheligovsky, V. Mean-field equations for weakly nonlinear two-scale
perturbations of forced hydromagnetic convection in a rotating layer,
{\it Geophys.~Astrophys.~Fluid Dyn.}, {\bf 102}, 2008, 489--540.

\mi Zheligovsky, V.A., Podvigina, O.M. and Frisch U. Dynamo effect
in parity-invariant flow with large and moderate separation of scales,
{\it Geophys. Astrophys. Fluid Dyn.}, {\bf 95}, 2001, 227--268.

\pagebreak
\mi{\bf Appendix. The hierarchy of equations for weakly nonlinear perturbations}

\mi
The following equations arise at order $\varepsilon^n$ after substitution
of the series \rf{pertser} for the perturbation into \xrf{eq5om}{eq5p3} and
expansion in power series, for the perturbed CHM state $\bOm,{\bf V,H},\Theta$
independent of~$\varepsilon$:
$${\cal L}^\omega(\bom_n,{\bf v}_n,{\bf h}_n,\theta_n)-{\partial\bom_{n-2}\over\partial T}
+\nu\lpar 2(\nabla_{\bf x}\cdot\nabla_{\bf X})\lb\bom_{n-1}\rb_v+\nabla^2_{\bf X}\bom_{n-2}\rpar$$
$$+\nabla_{\bf X}\times\lpar{\bf V}\times\bom_{n-1}+{\bf v}_{n-1}\times\bOm
-{\bf H}\times(\nabla_{\bf X}\times{\bf h}_{n-2}+\nabla_{\bf x}\times{\bf h}_{n-1})
-{\bf h}_{n-1}\times(\nabla_{\bf x}\times{\bf H})$$
$$+\sum_{k=0}^{n-2}({\bf v}_k\times\bom_{n-2-k}
-{\bf h}_k\times(\nabla_{\bf x}\times{\bf h}_{n-2-k}+\nabla_{\bf X}\times{\bf h}_{n-3-k}))\rpar$$
$$+\nabla_{\bf x}\times\lpar\sum_{k=0}^{n-1}({\bf v}_k\times\bom_{n-1-k}
-{\bf h}_k\times(\nabla_{\bf x}\times{\bf h}_{n-1-k}
+\nabla_{\bf X}\times{\bf h}_{n-2-k}))$$
$$-{\bf H}\times(\nabla_{\bf X}\times{\bf h}_{n-1})\rpar
+\beta\nabla_{\bf X}\theta_{n-1}\times{\bf e}_3={\bf 0},\eqno(\rm A.1)$$

$${\cal L}^h({\bf v}_n,{\bf h}_n)-{\partial{\bf h}_{n-2}\over\partial T}
+\eta\lpar 2(\nabla_{\bf x}\cdot\nabla_{\bf X})\lb{\bf h}_{n-1}\rb_h+\nabla^2_{\bf X}{\bf h}_{n-2}\rpar$$
$$+\nabla_{\bf X}\times\lpar{\bf v}_{n-1}\times{\bf H}+{\bf V}\times{\bf h}_{n-1}+\sum_{k=0}^{n-2}{\bf v}_k\times{\bf h}_{n-2-k}\rpar
+\nabla_{\bf x}\times\sum_{k=0}^{n-1}{\bf v}_k\times{\bf h}_{n-1-k}={\bf 0},\eqno(\rm A.2)$$

$${\cal L}^\theta({\bf v}_n,\theta_n)-{\partial\theta_{n-2}\over\partial T}
+\kappa\lpar 2(\nabla_{\bf x}\cdot\nabla_{\bf X})\theta_{n-1}+\nabla^2_{\bf X}\theta_{n-2}\rpar$$
$$-({\bf V}\cdot\nabla_{\bf X})\theta_{n-1}
-\sum_{k=0}^{n-1}({\bf v}_k\cdot\nabla_{\bf x})\theta_{n-1-k}
-\sum_{k=0}^{n-2}({\bf v}_k\cdot\nabla_{\bf X})\theta_{n-2-k}=0.\eqno(\rm A.3)$$

Averaging the vertical component of (A.1) in fast spatial variables, taking
into account relations \rf{eq17p2}, \rf{eq17p3} and \rf{eqomega} (valid for any
index $n$), \rf{eq11p1} and the boundary conditions for $\bf V$, $\bf H$,
${\bf v}_i$ and ${\bf h}_i$, we obtain
$${\partial\la\bom_n\ra_v\over\partial t}+{\partial\la\bom_{n-2}\ra_v\over\partial T}=
\nu\nabla^2_{\bf X}\la\bom_{n-2}\ra_v$$
$$+\,\nabla_{\bf X}\times\left\la{\bf V}\times(\nabla_{\bf X}\times{\bf v}_{n-2})
-{\bf V}\nabla_{\bf X}\cdot{\bf v}_{n-2}-{\bf H}\times(\nabla_{\bf X}\times{\bf h}_{n-2})
+{\bf H}\nabla_{\bf X}\!\cdot\!{\bf h}_{n-2}\phantom{|^|_|}\hskip-9pt\right.$$
$$+\sum_{k=0}^{n-3}\left.\!\!\lpar\!{\bf v}_k\!\times\!(\nabla_{\bf X}\!\times\!{\bf v}_{n-3-k})
\!-\!{\bf v}_k\nabla_{\bf X}\!\cdot\!{\bf v}_{n-3-k}
\!-\!{\bf h}_k\!\times\!(\nabla_{\bf X}\!\times\!{\bf h}_{n-3-k})
\!+\!{\bf h}_k\nabla_{\bf X}\!\cdot\!{\bf h}_{n-3-k}\rpar\!\right\ra_h\!\!.\eqno(\rm A.4)$$
\end{document}